\documentclass[10pt,twocolumn,letterpaper]{article}

\usepackage{cvpr}
\usepackage{times}
\usepackage{epsfig}
\usepackage{graphicx}
\usepackage{amsmath}
\usepackage{amssymb}

\usepackage{color}
\usepackage{booktabs}
\usepackage{threeparttable}
\usepackage{setspace}
\usepackage{multirow}
\usepackage{rotating}
\usepackage{array}
\usepackage{subcaption}
\makeatletter
\@namedef{ver@everyshi.sty}{}
\makeatother
\usepackage{tikz}
\usepackage{pgfplots}
\usetikzlibrary{shapes, arrows, arrows.meta, automata, positioning}
\usetikzlibrary{knots, calc}
\newcommand{\PreserveBackslash}[1]{\let\temp=\\#1\let\\=\temp}
\newcolumntype{C}[1]{>{\PreserveBackslash\centering}p{#1}}
\newcolumntype{R}[1]{>{\PreserveBackslash\raggedleft}p{#1}}
\newcolumntype{L}[1]{>{\PreserveBackslash\raggedright}p{#1}}
\graphicspath{{figures/}}

\usepackage[pagebackref=true,breaklinks=true,letterpaper=true,colorlinks,bookmarks=false]{hyperref}

\cvprfinalcopy 

\def\cvprPaperID{****} 

\ifcvprfinal\fi
\begin{document}

\title{NTIRE 2020 Challenge on Spectral Reconstruction from an RGB Image}

\author{Boaz Arad
\and Radu Timofte
\and Ohad Ben-Shahar
\and Yi-Tun Lin 
\and Graham Finlayson
\and Shai Givati
\and Jiaojiao Li
\and Chaoxiong Wu
\and Rui Song
\and Yunsong Li
\and Fei Liu
\and Zhiqiang Lang
\and Wei Wei
\and Lei Zhang
\and Jiangtao Nie
\and Yuzhi Zhao
\and Lai-Man Po
\and Qiong Yan
\and Wei Liu
\and Tingyu Lin
\and Youngjung Kim
\and Changyeop Shin
\and Kyeongha Rho
\and Sungho Kim  
\and Zhiyu ZHU
\and Junhui HOU
\and He Sun
\and Jinchang Ren
\and Zhenyu Fang
\and Yijun Yan
\and Hao Peng
\and Xiaomei Chen
\and Jie Zhao
\and Tarek Stiebel
\and Simon Koppers
\and Dorit Merhof
\and Honey Gupta
\and Kaushik Mitra
\and Biebele Joslyn Fubara
\and Mohamed Sedky
\and Dave Dyke
\and Atmadeep Banerjee
\and Akash Palrecha
\and Sabarinathan
\and K Uma
\and D Synthiya Vinothini
\and B Sathya Bama
\and S M Md Mansoor Roomi
}

\maketitle

\begin{abstract}
This paper reviews the second challenge on spectral reconstruction from RGB images, \ie, the recovery of whole-scene hyperspectral (HS) information from a 3-channel RGB image. 
As in the previous challenge, two tracks were provided: (i) a ``Clean'' track where HS images are estimated from noise-free RGBs, the RGB images are themselves calculated numerically using the ground-truth HS images and supplied spectral sensitivity functions (ii) a ``Real World'' track, simulating capture by an uncalibrated and unknown camera, where the HS images are recovered from noisy JPEG-compressed RGB images. 
A new, larger-than-ever, natural hyperspectral image data set is presented, containing a total of 510 HS images.
The Clean and Real World tracks had 103 and 78 registered participants respectively, with 14 teams competing in the final testing phase.
A description of the proposed methods, alongside their challenge scores and an extensive evaluation of top performing methods is also provided. They gauge the state-of-the-art in spectral reconstruction from an RGB image.
\end{abstract}

\section{Introduction}

{\let\thefootnote\relax\footnotetext{B. Arad (boazar@post.bgu.ac.il, Voyage81 \& Ben-Gurion University of the Negev), R. Timofte, O. Ben-Shahar, Y.-T. Lin, G. Finlayson, S. Givati are the NTIRE 2020 challenge organizers, while the other authors participated in the challenge. 
\\Appendix~\ref{sec:appendix} contains the authors' teams and affiliations.
\\NTIRE 2020 webpage:\\~\url{https://data.vision.ee.ethz.ch/cvl/ntire20/}}}

While conventional color cameras record scene spectral radiance integrated three spectral bands (red, green, and blue), hyperspectral imaging systems (HISs) can record the actual scene spectra  over a large set of narrow spectral bands~\cite{chang2007hyperspectral}. However, the rich, spectral, information provided by HISs comes with significant additional capture complexity: most common HISs rely on either spatial or spectral scanning (\eg push-broom or variable-filter systems) and hence are unsuitable for real-time operation. Moreover, hyperspectral capture often requires a longer capture time and this means it is difficult to measure information from scenes with moving content. Although, recent advances in ``Snapshot'' HISs have continued to bridge the gap towards real-time spectral image acquisition - \eg  Mosaic~\cite{tanriverdi2019dual, geelen2014compact, gonzalez2016novel} and light-field~\cite{beletkaia2020more} based snapshot HISs can capture images at video-rates -  these technologies record images with reduced spatial and spectral resolution. To date, both scanning and snapshot HISs remain prohibitively expensive for consumer grade use (``low-cost'' HISs are often in the \$10K-\$100K range). 

Due to these  drawbacks of HISs, there has been a lot of research and industrial interest in developing methods for recovering spectra from the images of low cost and ubiquitous RGB cameras. Early work on RGB spectral recovery images leveraged sparse coding methods to recover HS data~\cite{arad2016sparse,robles2015single,aeschbacher2017defense,timofte2014a+}. In recent years, neural-net based methods have become more common~\cite{galliani2017learned,can2018efficient,kaya2019towards,arad2018ntire,Shoeiby_2018_ECCVW,Shoeiby_2018_ECCV_Workshops,Lahoud_2018_ECCV_Workshops,TIC-RC_HSCNN+}, with leading methods from the NTIRE 2018 spectral recovery challenge~\cite{arad2018ntire,TIC-RC_HSCNN+} as well as more recent works~\cite{zhang2019hyperspectral,miao2019lambda,kaya2019towards,fu2020hyperspectral} adopting this approach. This transition to neural-net based methods highlights the need for larger data sets -  both to facilitate improved training as well as improved evaluation. The latter consideration is crucial as neural-nets are prone to ``overfitting'' on small data sets and thus their test scores may not generalize well to real-world applications.

The inherent difficulty in evaluation neural-net based solutions was recently highlighted by Yi-Tun and Finlayson~\cite{lin2019exposure} which evaluated top performing solutions from the NTIRE 2018 challenge under variable illumination conditions. Surprisingly, simply varying the brightness of input images (simulating longer/shorter camera exposures of the same scene) degraded the performance of neural net based methods to the point that they were outperformed by sparse-coding based methods (because the evaluated sparse-coding based methods were exposure invariant). Concomitantly, in Section~\ref{sec:res} we present an extended evaluation of top performing methods - which includes the variable exposure test - to  more thoroughly review the algorithms' performance.

Following NTIRE 2018, two potential experimental evaluation issues were identified and thus addressed here. First,
the top performing methods in the NTIRE 2018 challenge obtained a percentage recovery error of about 1\% mean relative absolute error (MRAE; \cf Section~\ref{sec:tracks} and Eq.~\ref{eq:mrae}), indicating that evaluation data may need better ``dynamic range'' beyond the one currently provided  by the BGU HS Dataset~\cite{arad2016sparse} or that evaluation should extend beyond the spectral quantization levels of 31 bands currently in use. 
Second, it was found that the ranking of the algorithms in the previous challenge did not differ significantly between the clean and real world tracks, possibly indicating that the simulated ``real world'' camera did not add sufficient complexity relative to the clean track. To this end, the 2020 challenge presents a larger-than-ever data set nearly twice as large as the BGU HS data set (\cf Section~\ref{sec:dataset}) as well as an improved real world track where camera noise  is  incorporated as well(\cf Section~\ref{sec:tracks}).

\section{NTIRE 2020 Challenge}
The RGB to spectra recovery challenge~\cite{arad2020ntire} is one of the NTIRE 2020  challenges. The other challenges are: deblurring~\cite{nah2020ntire}, nonhomogeneous dehazing~\cite{ancuti2020ntire}, perceptual extreme super-resolution~\cite{zhang2020ntire}, video quality mapping~\cite{fuoli2020ntire}, real image denoising~\cite{abdelhamed2020ntire}, real-world super-resolution~\cite{lugmayr2020ntire} and demoireing~\cite{yuan2020demoireing}. 

As in the NTIRE 2018 Spectral Recovery Challenge~\cite{arad2018ntire}, the objectives of the NTIRE 2020 Challenge on Spectral Reconstruction are: (i) gauging and improving the state-of-the art in HS reconstruction from RGB images; (ii) comparing the different spectral recovery approaches; (iii) further expanding the amount of natural HS images available to the research community. Importantly, the 2020 challenge introduces not only a new and improved data set, but also an extended evaluation which attempts to gauge the expected performance of proposed methods beyond the scope of the challenge's test images. 

\subsection{ARAD HS Dataset}
\label{sec:dataset}
\begin{figure}
	\centering
	\includegraphics[width=\linewidth]{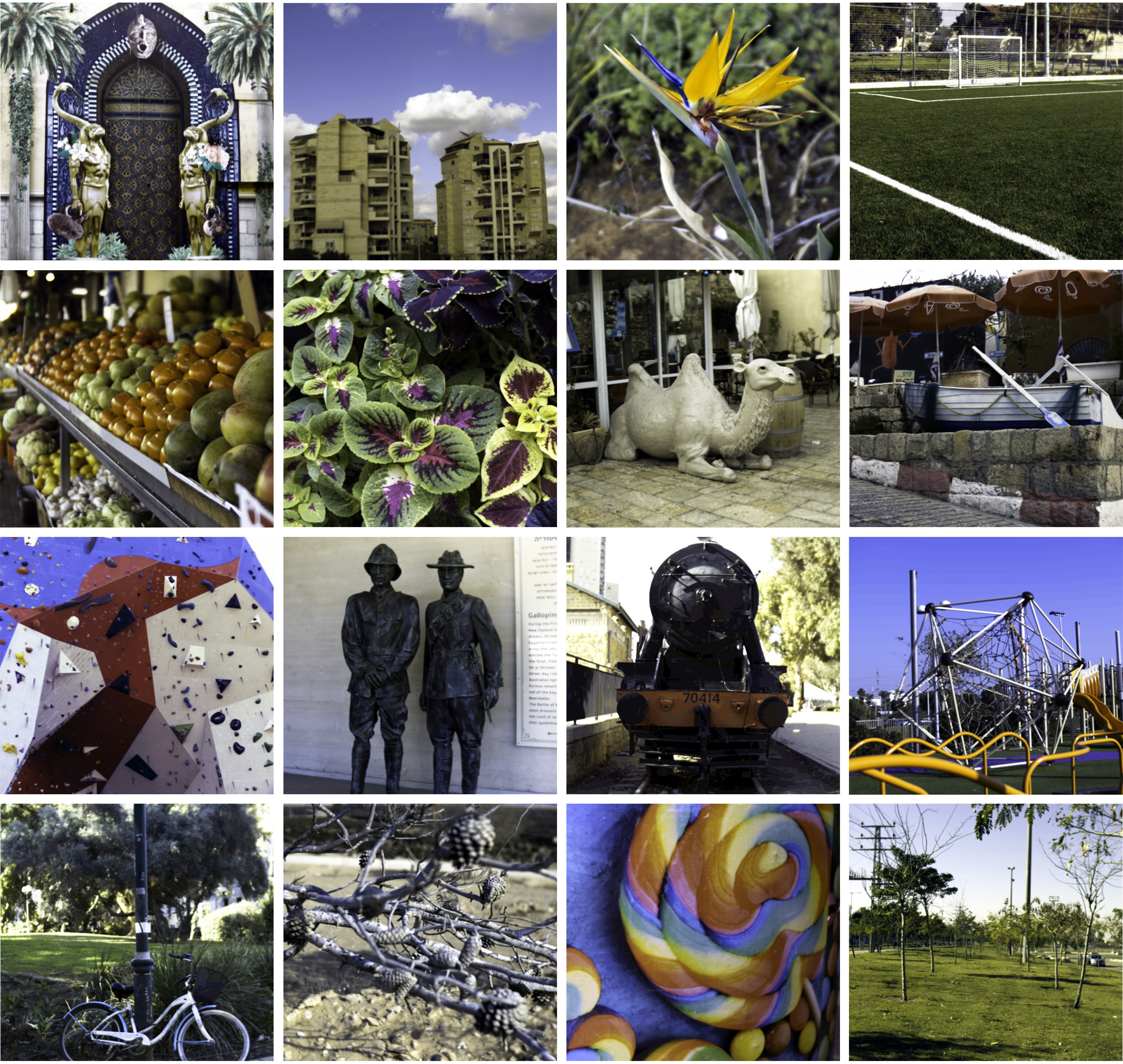}
	\caption{Sample images from the ARAD HS data set, note the variety of scene types (color and brightness have been manually adjusted for display purposes).}
	\label{arad_hs_examples}
\end{figure}

The NTIRE 2020 spectral reconstruction challenge provided a new, larger-than-ever, natural hyperspectral image data set. The data set included a total of 510 images: 450 training images, 30 validation images, and 30 test images. The training and test images were released during the challenge, while test images remain confidential to facilitate blind evaluation of future works. Figure~\ref{arad_hs_examples} includes a set of sample images from the data set.

The ARAD data set was collected with a Specim IQ mobile hyperspectral camera. The Specim IQ camera is a stand-alone, battery-powered, push-broom spectral imaging system, the size of a conventional SLR camera (207 $\times$ 91 $\times$ 74 mm) which can operate independently without the need for an external power source or computer controller. The use of such a compact, mobile system facilitated collection of an extremely diverse data set with a large variety of scenes and subjects. 

In addition to the ARAD data set, participants were invited to use the previously published BGU HS data set~\cite{arad2016sparse, arad2018ntire} as well to obtain a total of 706 training images.

\subsubsection{Radiometric Calibration}

The Specim IQ camera provides RAW $512\times 512$px images with 204 spectral bands in the 400-1000nm range. For the purpose of this challenge, manufacturer-supplied radiometric calibration has been applied to the RAW images, and the images have been resampled to 31 spectral bands in the visual range (400-700nm). Both RAW and radiometrically calibrated images have been made available to researchers.

The radiometric calibration corrects for measurement biases introduced by the camera systems CMOS sensor, converting the recorded RAW per channel intensity data to accurate spectral measurements. ``Lines'' (image columns) with excessive interference are also removed by this process, resulting in a $482\times 512$px image, resampled to 31 bands from 400nm to 700nm with a 10nm step.

\subsection{Tracks}
\label{sec:tracks}

As in the previous iteration of this challenge~\cite{arad2018ntire}, the NTIRE 2020 Spectral Recovery Challenge had two tracks, a ``clean'' and a ``real world'' track. While the clean track was similar to that of the previous challenge (NITRE 2018), the real world track was substantially updated to provide a more accurate simulation of physical camera systems.


\noindent\textbf{Track 1: ``Real World''} simulates the recovery of spectral information from an unknown, uncalibrated camera. Participants were provided with 8-bit color images in compressed JPEG format created by applying the following procedure to spectral images:
\begin{enumerate}
    \item Applying a real-world camera response function to a spectral image.
    \item Subsampling the resulting 3 channel image to produce an RGGB Bayer mosaic image.
    \item Adding simulated camera noise (Poisson shot noise and normally distributed dark noise) to the mosaic image.
    \item Applying a demosaicing algorithm from the OpenCV~\cite{opencv_library} library to produce a three-channel RGB image.
    \item Storing the image in compressed JPEG format. 
\end{enumerate}
The camera response and noise parameters used in the above procedure were kept confidential from challenge participants and shall remain confidential to facilitate equal ground comparisons of future works to the challenge results below.

Challenge participants were provided with code~\cite{NTIRE2020_spectral_code} (publicly available on the GitHub platform) used to generate both clean and real world track images.

\noindent \textbf{Competitions} competitions were hosted on the CodaLab platform~\footnote{\url{https://codalab.org/}}, with a separate competition for each track. After registration, participants were able to access data and submit results for automatic evaluation on the competition test server. Due to constraints of the CodaLab platform, the validation and test set have been reduced to 10 images each track (for a total of 10 validation images and 20 test images).

\noindent\textbf{Challenge phases} The challenge had two phases:
\begin{enumerate}
    \item \textbf{Development:} participants were provided with ground truth training hyperspectral/RGB image pairs for both tracks (450 image pairs for each track) as well as 10 RGB images for validation. A test server was made available to participant, allowing them to upload their results and receive an online evaluation score.
    \item \textbf{Testing:} ground truth spectral validation images were released, along with final test images for each track. Participants were invited to upload their final solutions to the test server, and results were kept confidential until the challenge concluded.  
\end{enumerate}

\noindent\textbf{Evaluation protocol} .
As in the 2018 competition~\cite{arad2018ntire}, Mean Relative Absolute Error (MRAE) computed between the submitted reconstruction results and the ground truth images was selected as the quantitative measure for the competition. Root Mean Square Error (RMSE) was reported as well, but not used to rank results. MRAE and RMSE are computed as follows:
\begin{equation}
MRAE = \frac{\sum_{i,c}\frac{|P_{gt_{i_c}}-P_{rec_{i_c}}|}{P_{gt_{i_c}}}}{|P_{gt}|},
\label{eq:mrae}
\end{equation}
\begin{equation}
RMSE =  \sqrt{\frac{\sum_{i,c} \left(P_{gt_{i_c}}-P_{rec_{i_c}}\right)^2 }{|P_{gt}|}},
\label{eq:rmse}
\end{equation}
where $P_{gt_{i_c}}$ and $P_{rec_{i_c}}$ denote the value of the $c$ spectral channel of the $i$-th pixel in the ground truth and the reconstructed image, respectively, and $|P_{gt}|$ is the size of the ground truth image (pixel count $\times$ number of spectral channels).

\section{Challenge Results}
\label{sec:res}
Submissions provided by challenge participants were evaluated against confidential ground-truth HS test-set images using the metrics described in Section~\ref{sec:tracks} (\cf Eq. \eqref{eq:mrae}, \eqref{eq:rmse}). The results of the evaluations are shown in Table~\ref{tab:challenge_results}. Self-reported computational requirements and additional implementation details for submitted methods are reported in Table~\ref{tab:factsheets}. The top performing method in the clean track (IPIC\_SSR) achieved a MRAE of 0.0301 and a RMSE of 0.0129.  The top performing method in the real world track (OrangeCat) achieved a MRAE of 0.0620 and a RMSE of 0.0192. For additional gains in accuracy top methods employed model ensemble and self-ensemble strategies~\cite{timofte2016seven}. All submitted solutions relied on recent-generation (and often state-of-the-art) GPUs for computation. Despite the use of powerful hardware, most solutions required at least 0.5 seconds to process a $\sim0.25$ mega-pixel (mp) image. The best placed solution that could recover the HSI in  less than 0.5 seconds per-image (LFB) was ranked  8th in the clean track and 5th in the real world track. To achieve recovery in less than 0.1 seconds per-image (StaffsCVL) we needed to go down the ranked list respectively to 10th and 8th position for the clean and real world tracks.  

In addition to the primary evaluation metrics, five additional auxiliary metrics were used to explore the stability and extrapolability of solutions proposed by participants. These metrics are described in the following sub-sections and were applied to the top-performing submissions. First,  \textbf{"out-of-scope"} images which differ significantly from the training data were considered. Second, \textbf{"shuffled"} images where large-scale spatial features are broken down to randomly ordered $4\times4$ patches were used gauge the robustness of methods to unseen conditions and/or spatial features. Then, third, test image \textbf{brightness} is varied to assess methods' stability under varying illumination intensity. In a fourth test, \textbf{weighted} scoring is applied to accurately represent performance over spectral signatures which have lower abundance in the test data. Finally, recovered HS images are projected back to RGB space to examine the \textbf{physical consistency} of results - do the recovered HS images produce RGB projections which are similar to the query images? This last test is interesting because if a method does not meet this criterion, regardless of the MRAE, it must be recovering incorrect spectra.

\begin{table*}[th!]
\centering
\begin{tabular}{ll||c|c||c|c}
     		    &             		& \multicolumn{2}{c||}{Track 1: Clean} 			& \multicolumn{2}{c}{Track 2: Real World} \\
Team 		                                 &  Username        	     & MRAE         		& RMSE        		& MRAE                  & RMSE           \\ \hline\hline
IPIC\_SSR~\cite{Jiaojiao_2020_CVPR_Workshops}&	Deep-imagelab            & 0.03010	$_{(1)}$	& 0.01293	        & 0.06216	$_{(3)}$	& 0.01991  \\
MDISL-lab                                    &	ppplang	                 & 0.03075	$_{(2)}$	& 0.01268	        & 0.06212	$_{(2)}$	& 0.01946 \\
OrangeCat~\cite{Zhao2020_Hierarchical}       &	zyz987	                 & 0.03231	$_{(3)}$	& 0.01389	        & 0.06200	$_{(1)}$	& 0.01923 \\
AIDAR	                                     &  PARASITE	             & -	                & -	                & 0.06514   $_{(4)}$    & 0.02065 \\
VIPLab$^1$                                   &	ZHU\_zy	                 & 0.03475	$_{(4)}$	& 0.01475	        & -		                & - \\
TIC-RC                                       &	sunnyvick	             & 0.03516	$_{(5)}$	& 0.01567	        & 0.07032	$_{(7)}$	& 0.02191 \\
VIPLab$^2$                                   &	ninaqian	             & 0.03518	$_{(6)}$	& 0.01511	        & -		                & - \\
GD322                                        &	Hpeng	                 & 0.03601	$_{(7)}$	& 0.01695	        & 0.06780	$_{(6)}$	& 0.02071 \\
LFB                                          &	Tasti 	                 & 0.03633	$_{(8)}$	& 0.01690	        & 0.06732	$_{(5)}$	& 0.02124 \\
CI Lab                                       &	honeygupta	             & 0.03769	$_{(9)}$	& 0.01677	        & 0.07581	$_{(9)}$	& 0.02253 \\
StaffsCVL~\cite{Fubara_2020_CVPR_Workshops}  &	fubarabjs	             & 0.04401$_{(10)}$	    & 0.01978	        & 0.07141	$_{(8)}$	& 0.02173 \\
Pixxel AI                                    &	akashpalrecha	         & 0.04441	$_{(11)}$	& 0.01645	        & 0.09322	$_{(10)}$	& 0.02255 \\
Image Lab	                                 &  sabarinathan	         & 0.04577	$_{(12)}$	& 0.01595	        & -		                & - \\
\textit{disqualified}	                                 &  Achiever11	             & 0.17382	$_{(13)}$	& 0.04573	        & 0.16459 $_{(11)}$	    & 0.04743 
\end{tabular}
\caption{NTIRE 2020 Spectral Reconstruction Challenge results and final rankings on the ARAD HS test data.}
\label{tab:challenge_results}
\end{table*}

\begin{table*}[th!]
\centering
\small
\resizebox{\linewidth}{!}
{
\begin{tabular}{l||c|c|c|c|c|c|c|c}
                & \multicolumn{2}{c|}{Reported runtime per image (sec)} & & & & & \\
Team            & Clean     & Real World& Platform   & CPU                  & GPU                       & Training Time & Notes                                           &Ensemble/Fusion\\ \hline
\hline
IPIC\_SSR~\cite{Jiaojiao_2020_CVPR_Workshops} &	0.56	& 0.56	    & Pytorch	 & E5-2678              & 2x NVIDIA 2080Ti 11G      & 36 hours		& Self-ensemble used only for "Real World" track. &self-ensemble, model-ensemble\\
MDISL-lab	    &	16	    & 16	    & PyTorch	 &                      & NVIDIA 1080 Ti 12G	    & 48 hours		&                                                 & 10 model ensemble \\
OrangeCat~\cite{Zhao2020_Hierarchical}	    &	3.74	& 3.74	    & PyTorch	 &                      & 2x NVIDIA Titan Xp 12G	& 7 days		& 8-setting ensemble strategy for both tracks     & self-ensemble, model-ensemble\\
AIDAR	        &	-	    & 30		& PyTorch    &                      & NVIDIA Titan Xp 12G	    & 16 hours		&                           & self-ensemble, model-ensemble \\
VIPLab$^1$      &	\~{}1	& -			& PyTorch    &                      & 4x UNKNOWN	            & 12 hours		&                           &                                   \\
TIC-RC	        &	0.7	    & 0.7	    & PyTorch	 & Intel(R) Xeon(R) CPU & Tesla K80 12GB	        & 13.9 hours	&                           &	                                \\
VIPLab$^1$      &	\~{}1	& -			& PyTorch    &                      & 4x UNKNOWN	            & 12 hours		&                           &                                   \\
GD322	        &	1.35	& 1.35	    & PyTorch	 & E5-2680              & NVIDIA Titan Xp	        & 13 hours		&                           &                                   \\
LFB	            &	0.31	& 0.30	    & PyTorch    & Intel Xeon W-2133    & NVIDIA GTX 2080Ti			&   8 hrs       &                           &                                   \\
CI Lab	        &	0.4	    & 0.4	    & TensorFlow & Intel i9-9900X       & NVIDIA GTX 2080Ti	        & 36 hours		&                           &                                   \\
StaffsCVL~\cite{Fubara_2020_CVPR_Workshops}	    &	0.034	& 0.034	    & PyTorch    & Intel Core i5        & NVIDIA RTX 2080Ti         & 2.7 hours     &                           &                                   \\
Pixxel AI	    &	0.154	& 0.154			& FastAI     & Intel(R) Xeon(R)     & NVIDIA V100               & 8.6 hours     &                           &                                   \\
Image Lab	    &	0.69	& -			& Keras      & Intel Core i7        & NVIDIA GTX 1080			&               &                           &                                   \\
\textit{disqualified}	    &   3.75	& 3.75      & Keras      & Intel Core i7        & NVIDIA GTX 1080		    &  33 hours     &                           &                                   
\end{tabular}
}

\caption{Self-reported runtimes per image on the ARAD HS test data and additional implementation details.}
\label{tab:factsheets}
\end{table*}

\begin{table*}[ht!]
\centering
\resizebox{\linewidth}{!}
{
\begin{tabular}{l||c|c|c|c|c|c||c|c|c|c|c|c}
& \multicolumn{6}{c||}{Track 1: Clean (MRAE)} & \multicolumn{6}{c}{Track 2: Real World (MRAE)} \\
Team & Out-of-Scope & Spatial & Brightness$\times 0.5$ & Brightness$\times 2$ & Physical & Weighted & Out-of-Scope & Spatial & Brightness$\times 0.5$ & Brightness$\times 2$ & Physical & Weighted \\ \hline\hline
IPIC\_SSR~\cite{Jiaojiao_2020_CVPR_Workshops} &  0.08511 & 0.09580 & 0.03273 & 0.03969 & 0.00117 & 0.03746 & 0.12556 & 0.45796 & 0.08000 & 0.06598 & 0.03290 & 0.07944\\
MDISL-lab & 0.08076	& 0.07948 & 0.03562 & 0.03390 & 0.00053 & 0.03901 & 0.14005 & 0.21058 & 0.08203 & 0.06832 & 0.03563 & 0.07981\\
OrangeCat~\cite{Zhao2020_Hierarchical} & 0.09233 & 0.07670 & 0.04052 & 0.04419 & 0.00103 & 0.04169 & 0.13019 & 0.22689 & 0.08097 & 0.06784 & 0.03346 & 0.07945\\
\end{tabular}
}
\caption{Auxiliary test results of the top 3 models in each track. \textbf{Out-of-scope:} performance on images which differ significantly from the training data. \textbf{Spatial:} performance on images which were broken down to randomly ordered $4\times4$ patches. \textbf{Brightness ($\times0.5$, $\times2$):} performance over images where intensity was decreased/increased by $\times0.5/\times2$. \textbf{Physical:} correlation between RGB projection of recovered spectra and input RGB images. \textbf{Weighted:} accuracy over representative spectra samples without accounting for their abundance. }
\label{tab:additional_tests}
\end{table*}

\subsection{Performance on ``Out-of-Scope'' Images}
\begin{figure}
	\centering
	\includegraphics[width=0.95\linewidth]{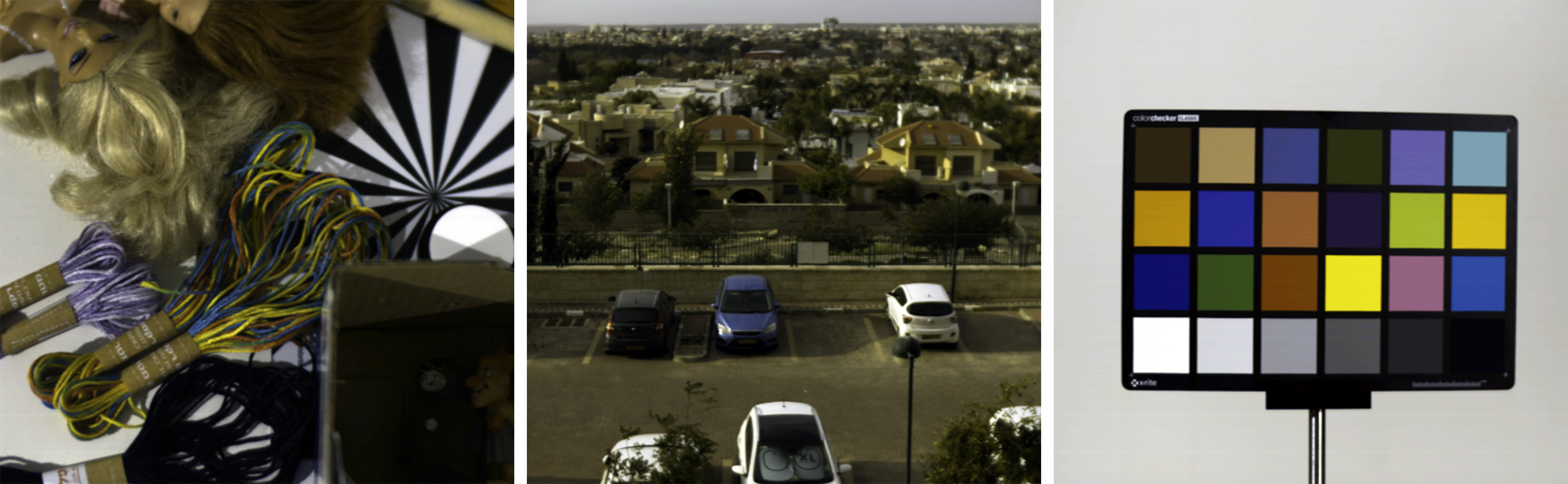}
	\caption{Sample ``out-of-scope`` images used to evaluate the proposed methods' generalization capabilities. Studio images (left, right) were recorded under halogen illumination. All images were images manually adjusted (color/brightness) for display.}
	\label{fig:out_of_scope}
\end{figure}
To study the generalizability of the proposed models, the top 3 models of each track are tested with 5 additionally images that were taken under drastically different settings, \eg objects in a studio, halogen lighting, scenes with rare viewing perspective, etc. Example images are given in Figure~\ref{fig:out_of_scope}.

The MRAE results of this study are given in the ``Out-of-Scope'' columns of Table~\ref{tab:additional_tests}. Perhaps unsurprisingly, the average MRAE error calculated for the out-of-scope images is more than \textit{doubled} the MRAE score of all top performers, but perhaps more interestingly - solution ranking for top performers is varied significantly from their ranking on the challenge test set. Indicating, perhaps, that a slightly lower performance on data similar to the training set might be acceptable if the method is to generalise to spectral images that are quite different. 

\subsection{Dependence on Spatial Features}

While most of the challenge participants exploit high-level information (\ie image content) by mapping large image patches, many pixel-based spectral reconstruction methods in the prior art have already shown efficacy to a certain extent, \eg \cite{nguyen2014training,arad2016sparse,lin2019exposure,aeschbacher2017defense}. 
The purpose of this study is to examine: to what degree the proposed models can retain their efficacy if spatial information of the test images is, by construction, much more limited.

Models are tested with the ``spatially shuffled'' test images: each $4\times 4$ patch in the original test images is randomly relocated. The MRAE results are given in the ``Spatial'' columns of Table~\ref{tab:additional_tests}. In the clean track, a significant degradation in performance can be seen and again - solution ranking for top performers is varied  from their ranking on the challenge test set. In the real world track, degradation is dramatic, to the point where recovered data is unlikely to be usable (MRAEs of 0.22-0.45). It can be surmised that all top-performing solutions rely heavily on spatial information to overcome camera noise and compression artifacts in the real world track. In the noiseless clean track, dependence on spatial features remains significant, but much reduced relative to the real world track.

\subsection{Dependence on Image Brightness}

The RGB images can be brighter or dimmer depending on the exposure setting of the camera (\eg shutter speed and aperture size) and/or the varying illumination intensity of the scene, which corresponds to linearly scaled ground-truth spectra. This means a linearly scaled hyperspectral image and its RGB counterpart is also a physically valid ground-truth pair. 
However, the best models in the 2018 competition~\cite{arad2018ntire} appear to perform poorly when the scene brightness changes~\cite{lin2019exposure}. 

In this year's challenge, the tests with two brightness modulations are included: half (HS images scaled down by a factor of 0.5) and double (scaled up by a factor of 2). The corresponding clean-track and real-world-track RGB images are simulated following the original methodology. The results are shown respectively in the ``Brightness$\times 0.5$'' and ``Brightness$\times 2$'' columns of Table~\ref{tab:additional_tests}. While varied exposure caused performance degradation in this years' top performers as well, the scale of this degradation is significantly reduced relative to the previous competition's top performers (MRAE degraded by 32\% at most \vs 1245\% at most for the 2018 top performer \cite{lin2019exposure}). 

\subsection{Physical Consistency of Results}

The hyperspectral and RGB images are physically related. Indeed, following a specified pipeline, RGB images can be accurately simulated from hyperspectral images (refer to section~\ref{sec:tracks}). The so-called \textit{physical consistency} asks the question: if the reconstructed hyperspectral images are applied with the original pipeline and re-generate the RGB images, how far off are these re-generated RGB images from the original ones?

The results are presented as the MRAE between the ground-truth and re-generated RGB images in the ``Physical'' columns of Table~\ref{tab:additional_tests}. Top performers presented relatively high consistency with images in the clean track, and slightly reduced consistency with images in the real world track. Reduced consistency in the latter is likely attributable to simulated camera noise and compression artifacts. However, although the RGB MRAE numbers are small we make two additional comments. First, assuming (approximately the following assumption holds) that a 1\% MRAE error correlates roughly with Just Noticeable Differences (1 JND is a concept from psychophysics where an observer can just see the difference between stimuli) an MRAE of 3\% correlates with a color difference of 3 which in turn correlates with perceived colors in images than can be seen to be different. Second, the MRAE hides the fact that the, for example, 95\% quantile error can be large ($>10$). This kind of error means that the recovered spectrum, when projected back to the RGB, results in a color which is instantly noticeable as different. 

Curiously, because the recovered spectra do not reproject to the same RGB, these spectra cannot be the correct answer (irrespective of any MRAE).

\subsection{Weighted Accuracy}

The spectral properties of the pixels representing the same material are expected to be similar. However, the abundance of one material in the scene does not indicate its importance. This study aims to provide a fair assessment across different materials in each scene. First, similar spectra are grouped into 1000 clusters. Then, the mean MRAE of each groups are calculated individually. Finally, the weighted MRAE is the mean of the groups' performances.

The results are provided in the ``Weighted'' columns of Table~\ref{tab:additional_tests}.

\section{Conclusions}
The NTIRE 2020 Challenge on Spectral Reconstruction from an RGB Image provides the most extensive evaluation to date of methods for spectral recovery from RGB images in terms of both participation and evaluation scope. Participants were provided with a larger-than-ever natural hyperspectral image data set and presented a wide variety of neural net based solutions to the task of spectral recovery from RGB images. Analysis of the proposed solutions revealed several intriguing areas for future exploration, namely: high-performance spectral recovery for video and/or edge devices, reducing dependence on spatial features, and increased robustness to unseen scenes. 

Top performing methods required \emph{at least} 0.5 seconds to process a $\sim0.25$mp image on two state-of-the-art GPUs, the fastest method required $\sim34$ms on a single state-of-the-art GPU. While the latter could claim processing at ``video rates'' (30fps), this would only hold true for $0.25$mp video on a GPU based platform. Extrapolating from this information, processing a single frame of 4K video (8.5mp) would require approximately 34 and 1.15 seconds on a single GPU for the most accurate and fastest method respectively. Processing on an edge device (\eg cellular phone) without a discrete GPU can be expected to take an order-of-magnitude longer. Future challenges may include an ``edge device'' track where solutions are scored on their computation requirements as well as their recovery performance.

All top performers were found to have a nontrivial dependency on spatial features when recovering spectral information from RGB images. The impact of this dependence becomes clear when one considers possible uses of recovered spectral information, for example: differentiating between similar objects based on their spectral reflectance (\eg real fruit vs. fake plastic fruit). For this reason future challenges may emphasize dependence on spatial features when scoring proposed methods and possibly include an application-based test metric as well. 

The tests on image brightness and physical consistency are interesting. For the same scene, exposure - how well the same physical object is lit - varies across the scene. But, despite this we would expect to recover the spectrum (up to a scaling factor) and this was found not to be the case for existing methods. The physical consistency test is interesting and surprising. All challenge methods do not find spectra consistent with the original RGB. Even though their MRAE may be small, these methods {\it must} recover the wrong answer. 

Finally, ``out-of-scoope'' image tests reveal that none of the top performers were able to robustly extrapolate to new settings. This indicates that while the training data set provided to participants is the largest of its kind, it can be further extended to cover additional settings. 
Namely indoor scenes and scenes under a larger variety of illumination conditions should be added to future data sets. The constantly increasing portability and ease-of-use of modern HISs is expected to facilitate the collection of larger and more varied data sets.


\section{Challenge Methods and Teams}


\subsection{IPIC\_SSR - Adaptive Weighted Attention Network with Camera Spectral Sensitivity Prior for Spectral Reconstruction from RGB Images\cite{Jiaojiao_2020_CVPR_Workshops}}
\begin{figure}
	\centering
	\includegraphics[width=0.95\linewidth]{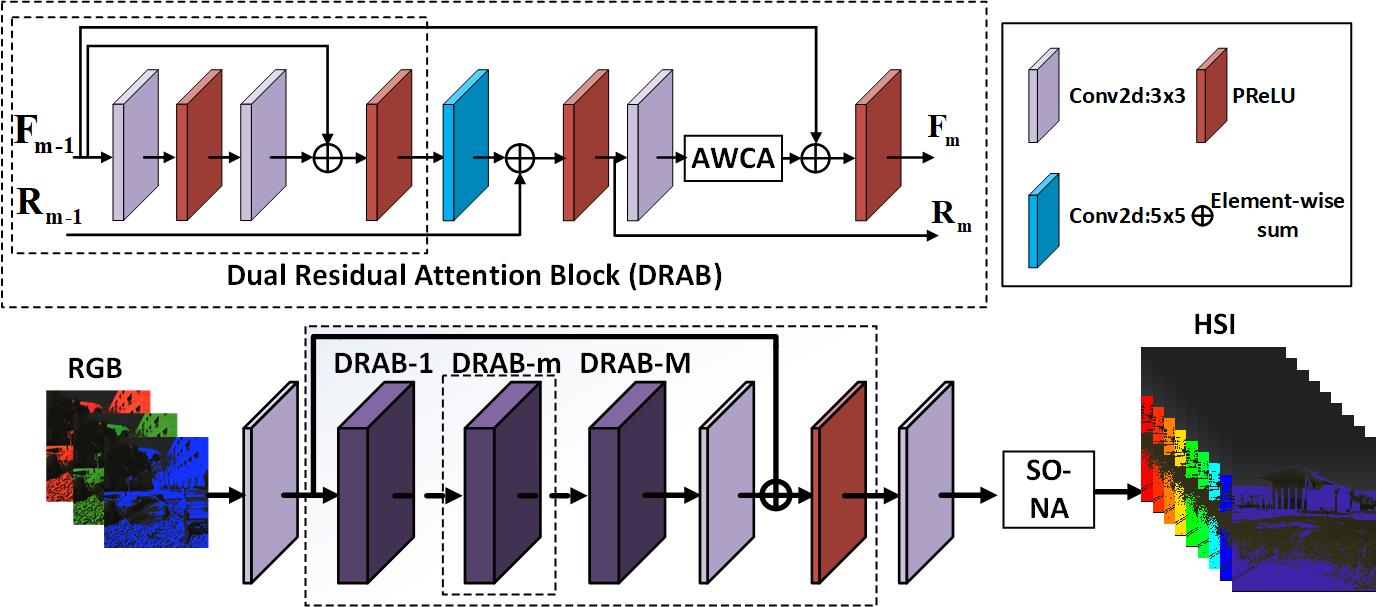}
	\caption{Network architecture of the IPIC\_SSR adaptive weighted attention network (AWAN).}
	\label{[IPIC_SSR]_[figure1]}
\end{figure}
\begin{figure}
	\centering
	\includegraphics[width=0.95\linewidth]{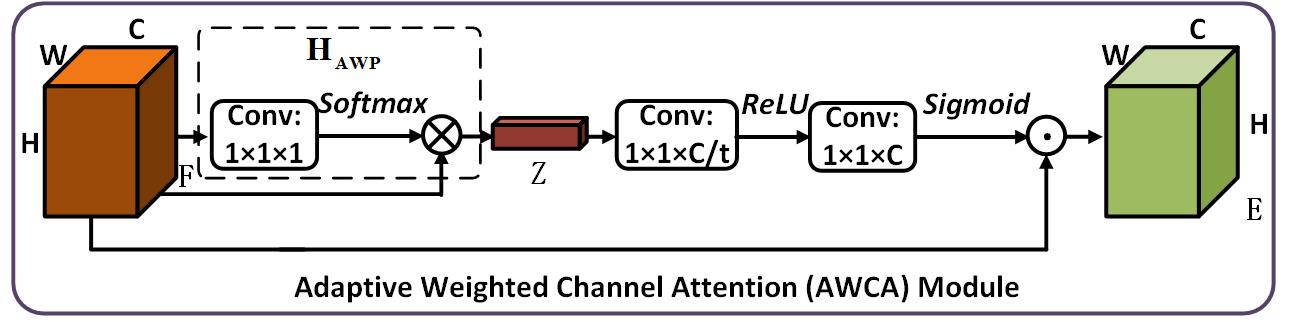}
	\caption{Diagram of adaptive weighted channel attention (AWCA) module. $\odot$ denotes element-wise multiplication.}
	\label{[IPIC_SSR]_[figure2]}
\end{figure}
\begin{figure}
	\centering
	\includegraphics[width=0.95\linewidth]{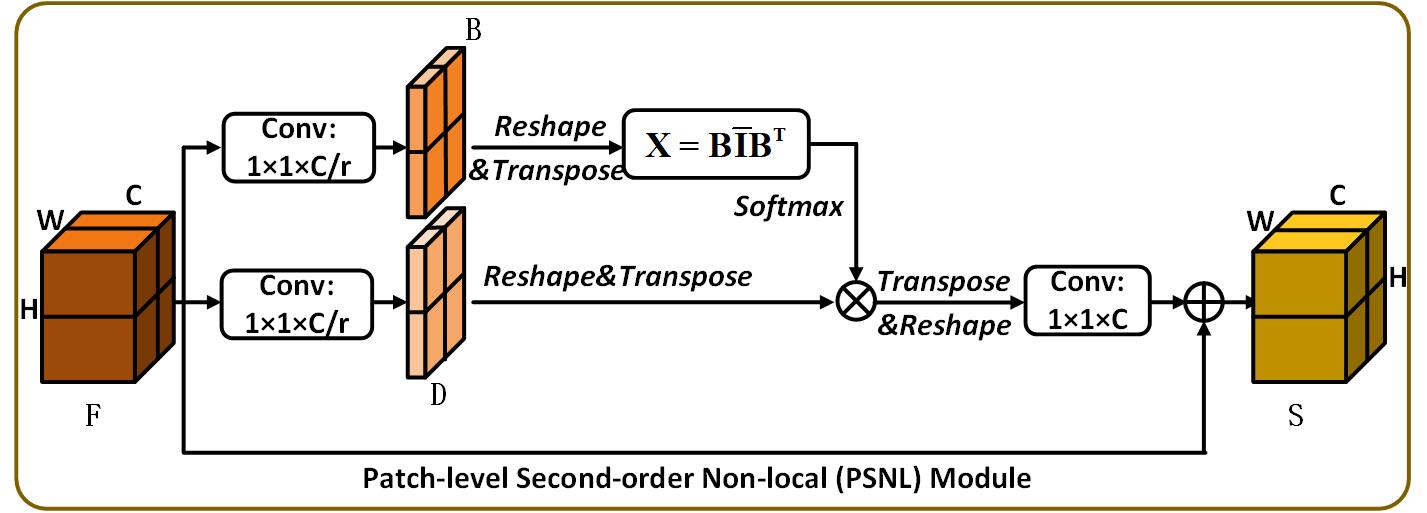}
	\caption{Diagram of patch-level second-order non-local (PSNL) module. $\otimes$ denotes matrix multiplication.}
	\label{[IPIC_SSR]_[figure3]}
\end{figure}

As shown in Figure~\ref{[IPIC_SSR]_[figure1]}, a novel deep adaptive weighted attention network (AWAN) is presented for spectral reconstruction from RGB images. Specifically, the backbone architecture of the AWAN network is constituted of 8 dual residual attention blocks (DRAB). Each DRAB consists of a traditional residual module and additional paired convolutional operations with a large (5 $\times$ 5) and small size (3 $\times$ 3) kernels, where the long and short skip connections to form the dual residual learning in the block. Typically, the output channel of each convolutional layer is set to 200. The adaptive weighted channel attention (AWCA) module (see in Figure \ref{[IPIC_SSR]_[figure2]}) embedded in the DRAB adaptively integrates channel-wise interdependencies. At the tail of the AWAN network, a patch-level second-order non-local (PSNL) module is employed to capture long-range spatial contextual information via second-order non-local operations. The diagram of the PSNL module is illustrated in Figure \ref{[IPIC_SSR]_[figure3]}).  
                      
Since the ``Clean'' track aims to recover hyperspectral images (HSIs) from the noise-free RGB images created by applying a known spectral response function to ground truth hyperspectral information, the camera spectral sensitivity (\ie spectral response function) prior is introduced to improve the quality of spectral reconstruction. Considering the fact that the reconstructed RGB can be calculated naturally through the super-resolved HSI, the final loss is a linear combination of the discrepancies of RGB images and the differences of HSIs
\begin{equation}
\label{IPIC_SSR_equa1}
l = l_{h} + \tau l_{r}
\end{equation}
where $\tau$ denotes the tradeoff parameter and is set to 10 empirically. Given the ground truth ${{\textbf{I}}_{HSI}}$ and the spectral super-resolved HSI ${{\textbf{I}}_{SSR}}$, the two loss functions are specifically defined as
\begin{equation}
\label{IPIC_SSR_equa2}
l_{h}=\frac{1}{N}\sum_{n=1}^{N}(|\textbf{I}_{HSI}^{(n)}-\textbf{I}_{SSR}^{(n)}|/\textbf{I}_{HSI}^{(n)}) 
\end{equation}
\begin{equation}
\label{IPIC_SSR_equa3}
l_{r}=\frac{1}{N}\sum_{n=1}^{N}(|\mathbf{\Phi}(\textbf{I}_{HSI}^{(n)})-\mathbf{\Phi}(\textbf{I}_{SSR}^{(n)})|) 
\end{equation}
where ${\textbf{I}_{HSI}^{(n)}}$ and  ${\textbf{I}_{SSR}^{(n)}}$ denote the ${n}$-th pixel value and ${\mathbf{\Phi}}$ is camera spectral sensitivity function. ${N}$ is the total number of pixels. However, the camera spectral sensitivity is unknown in the ``Real World'' track, thus the AWAN network is optimized by stochastic gradient descent algorithm with individual constraint $l_{h}$.

\subsubsection{Global Method Description}
\label{Global Method Description}
\noindent\textbf{Training}
During the training, ${64 \times 64}$ RGB and HSI sample pairs are cropped with a stride of 32 from the original dataset. The batch size of our model is 32 and the parameter optimization algorithm chooses Adam modification with ${\beta_1 = 0.9}$, ${\beta_2 = 0.99}$ and ${\epsilon = 10^{-8}}$. The reduction ratio $t$ value of the AWCA module is 16. The learning rate is initialized as 0.0001 the polynomial function is set as the decay policy with $power = 1.5$. The network training is stopped at 100 epochs. The proposed AWAN network has been implemented on the Pytorch framework and approximately 36 hours are required for training a network with 8 DRABs and output channel $=200$ on 2 NVIDIA 2080Ti GPUs .

\noindent\textbf{Testing} 
In our experiments, different spectral recovery ways are tried and compared with their scores in the validation sets of the two tracks. One way is to split the input images into small
overlapping patches, then average and stitch their outputs together on the GPU. The other is to feed the entire image to the AWAN network for inference on the CPU. Finally, the whole image is inputted into the network to fulfill the spectral recovery on the ``Clean'' track and at least 64G CPU is required for inference. The inference-time per image (CPU time) is 57.05s for both validation and test data. For the ``Real World'' track, the entire image is split into ${128 \times 128}$ overlapping patches with a stride of 64 and perform spectral reconstruction on an NVIDIA 2080Ti GPU with 11G memory. The AWAN network takes 0.56s per image (GPU time) for both validation and test data. By the way, in the ``Clean'' track, we can also achieve fast spectral reconstruction in the same way as the ``Real World'' track on the GPU, but the results will be slightly worse. 

\subsubsection{Ensembles and fusion strategies}
For the ``Clean'' track, four models are trained for model-ensemble strategy, including two models with 8 DRABs and 200 channels and two models with 20 DRABs and 128 channels. Different from the ``Clean'' track, for the ``Real World'' track, the self-ensemble method~\cite{timofte2016seven} is firstly adopted for single AWAN network. Concretely, the RGB input is flipped up/down to acquire a mirrored output. Then the mirrored output and the original
output are averaged into the target result. Also, three models with 8 DRABs and 200 channels and one model with 10 DRABs and 180 channels are trained for model-ensemble of AWAN network. Please refer to \cite{Jiaojiao_2020_CVPR_Workshops} for specific details.


\subsection{MDISL-lab - Improved Pixel-aware Deep Function-Mixture Network}

One fact is that the spectral of different pixels in a image vary widely. However, most existing Deep Convolutional Neural Networks (DCNNs) based Spectral Reconstruction (SR) methods treat all pixels in Hyper-Spectral Images equally and learn a universal mapping function, as shown in Figure.  Based on the observation, we present a pixel-aware deep function-mixture network for SR, which is flexible to pixel-wisely determine the receptive field size and the mapping function.\\

One fact is that the spectral of different pixels in a image vary widely. However, most existing Deep Convolutional Neural Networks (DCNNs) based Spectral Reconstruction (SR) methods treat all pixels in Hyper-Spectral Images equally and learn a universal mapping function, as shown in Figure.  Based on the observation and inspired by~\cite{MDISL-lab_zhang2019pixel}, a pixel-aware deep function-mixture network is presented for SR, which is flexible to pixel-wisely determine the receptive field size and the mapping function.\\

\begin{figure}
    \centering
    \includegraphics[height=1.4in, width=2.5in]{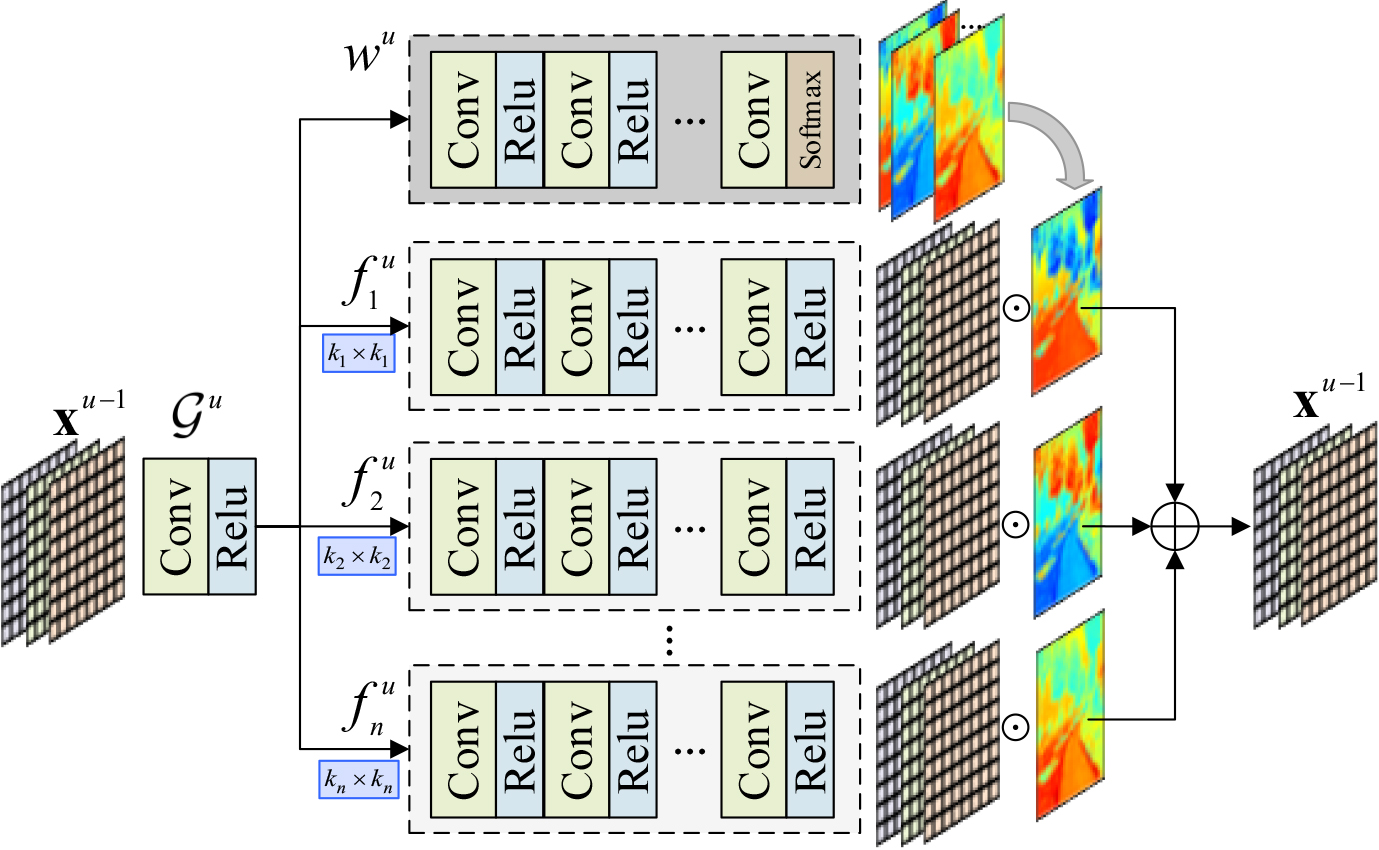}
    \caption{Architecture of the MDISL-lab function-mixture block.}
    \label{MDISL-lab_FMBlock}
\end{figure}

\begin{figure*}
    \centering
    \includegraphics[height=0.9in, width=5.3in]{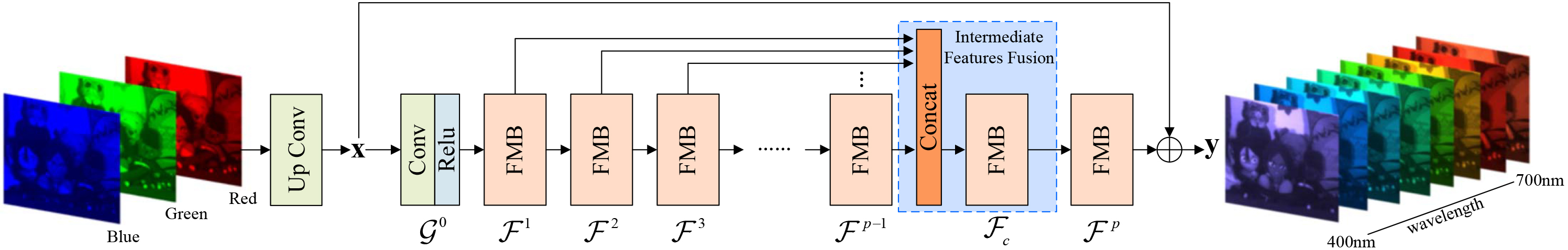}
    \caption{Architecture of the MDISL-lab pixel-aware deep function-mixture network. UpConv denotes a 3x3 convolution layer, which is used to increase the channels of input RGB image to the same as the output. FMB denotes the function-mixture block.}
    \label{MDISL-lab_FMNet}
\end{figure*}

It is worth noting that, in order to reduce the computational complexity, different receptive fields of different sizes are achieved by stacking multiple 3x3 convolution layers. To further improve the learning ability of the network, a SE module~\cite{MDISL-lab_hu2018squeeze} is placed after each branch and at the end of each module.

Specifically, a new module, termed the function-mixture (FM) block, is firstly developed. Each FM block consists of some parallel DCNN based subnets, among which one is termed the {\textit{mixing function}} and the remaining are termed {\textit{basis functions}}. The basis functions take different-sized receptive fields and learn distinct mapping schemes; while the mixture function generates pixel-wise weights to linearly mix the outputs of the basis functions, as shown in Figure~\ref{MDISL-lab_FMBlock}. In this way, the pixel-wise weights can determine a specific information flow for each pixel and consequently benefit the network to choose appropriate RGB context as well as the mapping function for spectrum recovery. Then, several such FM blocks are stacked to further improve the flexibility of the network in learning the pixel-wise mapping. Furthermore, to encourage feature reuse, the intermediate features generated by the FM blocks are fused in late stage. The overall architecture of proposed network is shown in figure~\ref{MDISL-lab_FMNet}.

\noindent\textbf{Training}
The paired spectral and RGB patches with a spatial size of 64x64 are cropped from the original images with a stride of 64. For data augmentation, horizontal flip and 90 degree rotation were  randomly performed. Teh model was trained by ADAM optimizer and the mini-batch size is set to 64. The initial learning rate was set to 3e-4 and halved at every 60,000 iterations for three times. The model was implemented  through PyTorch framework and  trained on a NIVIDIA 1080 Ti GPU and Intel(R) Xeon(R) CPU E5-2640.


\subsection{OrangeCat - Hierarchical Regression Network for Spectral Reconstruction from RGB Images~\cite{Zhao2020_Hierarchical}}

\begin{figure}[htbp]
\centering
\includegraphics[width=\linewidth]{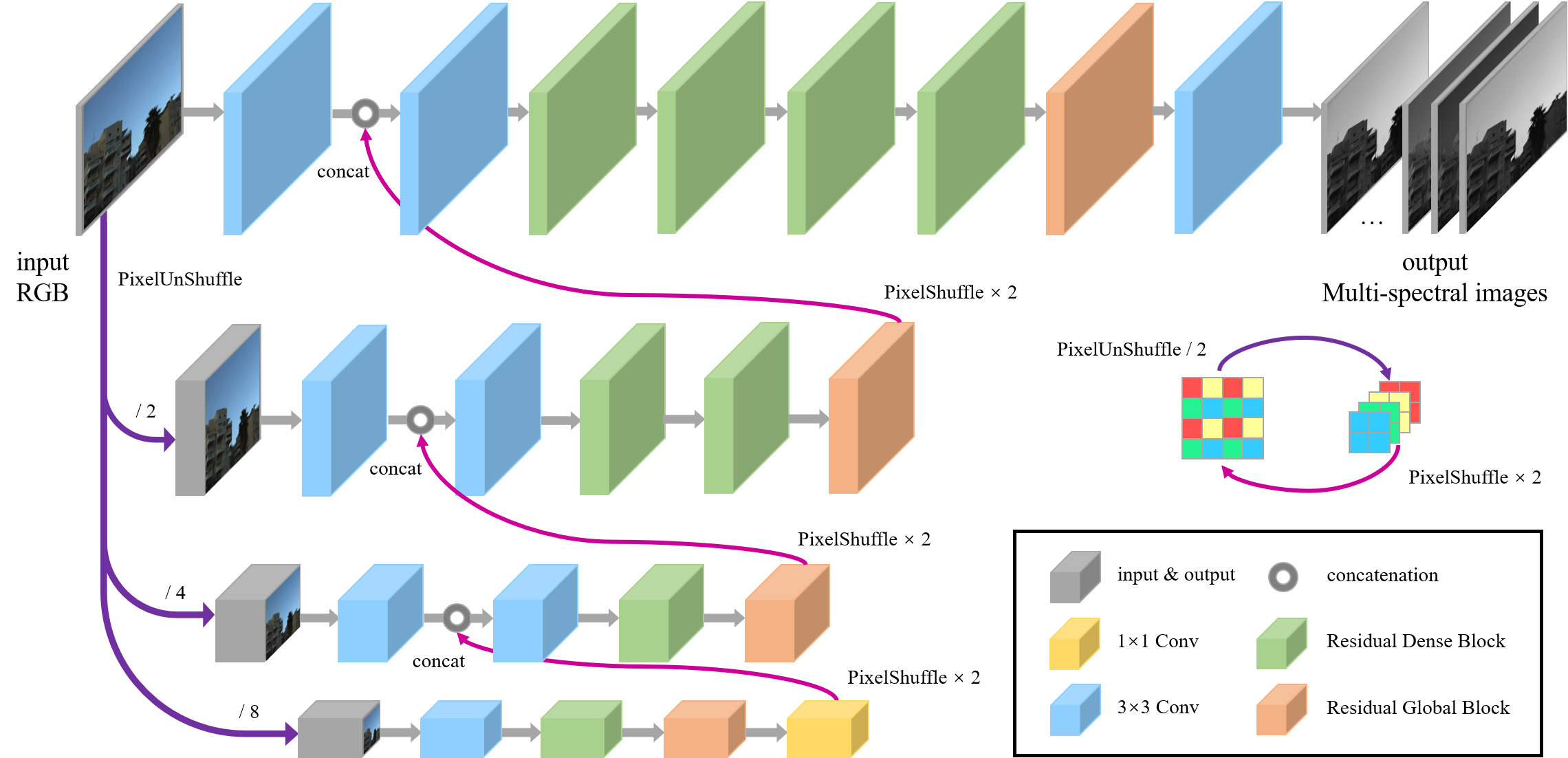}
\caption{Illustration of the HRNet architecture.}
\label{zyz_f1}
\end{figure}

Generally, we propose a 4-level hierarchical regression network (HRNet)~\cite{Zhao2020_Hierarchical} architecture for high-quality spectral reconstruction from RGB images, as shown in Figure \ref{zyz_f1}. The PixelShuffle layers~\cite{OrangeCat-PixelShuffle} are utilized to downsample the input to each level without adding parameters. Thus, the number of pixels of input is fixed while the spatial resolution decreases. Since PixelShuffle only reshapes feature maps and does not introduces interpolation operation, it allows HRNet to learn upsampling operation.

For each level, the process is decomposed to inter-level integration, artifacts reduction, and global feature extraction. The top level uses the most blocks to effectively integrate features and reduce artifacts thus produce high-quality spectral images. For inter-level learning, the output features of subordinate level are pixel shuffled, then concatenated to superior level, which uses an additional convolutional layer to unify the channels. In order to effectively reduce artifacts, we adopt a series of dense connection blocks~\cite{OrangeCat-DenseNet, OrangeCat-ResNet}, containing 5 convolutional layers and a residual. The residual global block~\cite{MDISL-lab_hu2018squeeze, OrangeCat-ResNet} with short-cut connection of input is used to extract different scales of features. In this block, each remote pixel is connected with other pixels to model the global attention due to MLP layers. The illustration of these blocks are in Figure 2.

\begin{figure}[htbp]
\centering
\includegraphics[width=\linewidth]{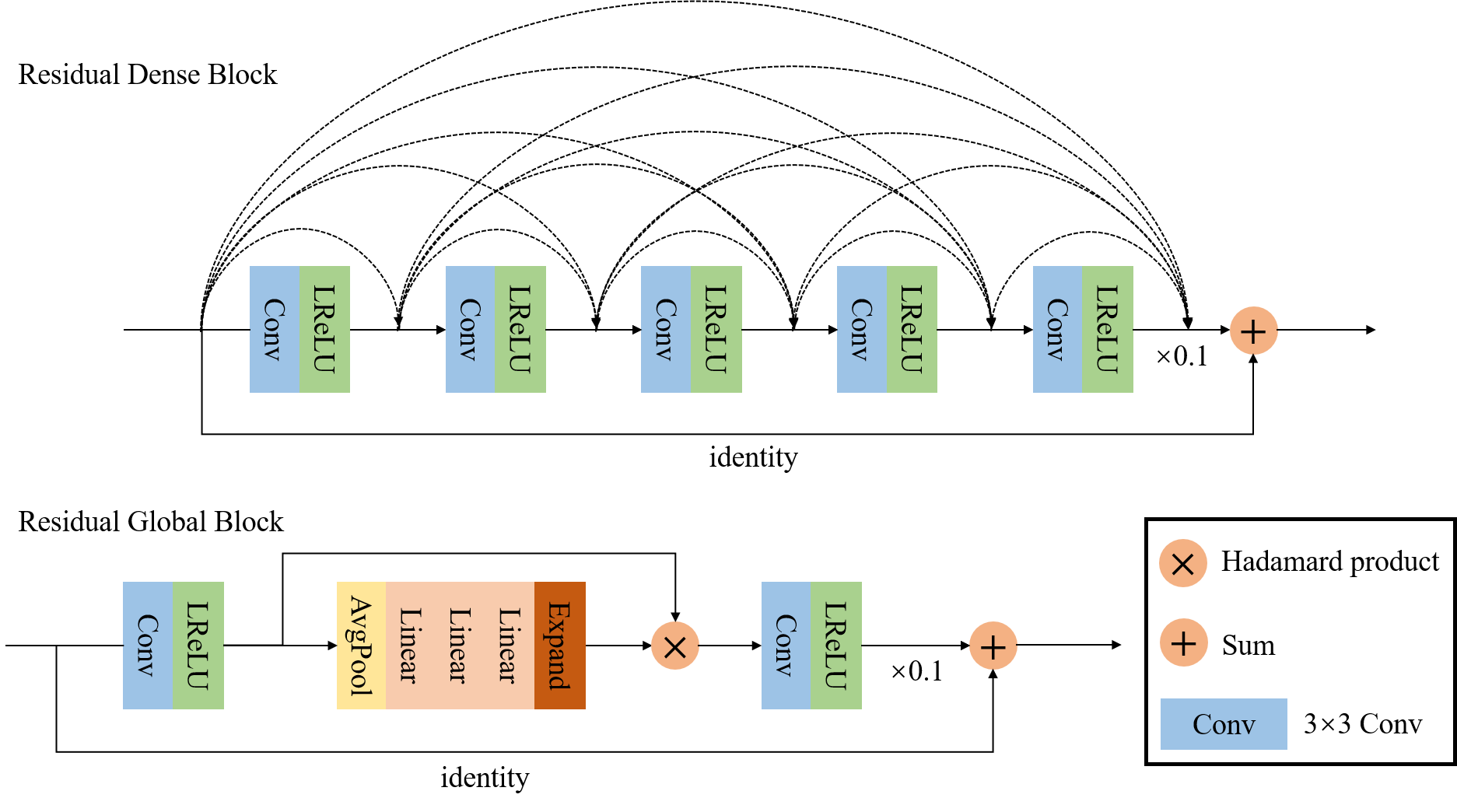}
\caption{Illustration of the residual dense block (RDB) and residual global block (RGB) architectures.}
\label{zyz_f2}
\end{figure}

Since the features are most compact in bottom level, there is a $1 \times 1$ convolutional layer attached to the last of bottom level in order to enhance tone mapping by weighting all channels. The two mid levels process features at different scales. Moreover, the top level uses the most blocks to effectively integrate features and reduce artifacts thus produce high-quality spectral images. The illustration of these blocks are in Figure \ref{zyz_f2}.

\noindent\textbf{Training} OnlyL1 loss is used duriung the training process, which is a PSNR-oriented optimization for the system. The L1 loss is defined as:

\begin{equation}
L_{1} = \mathbb{E}[||G(x) - y||_1],
\end{equation}
where $x$ and $y$ are input and output, respectively. The $G(*)$ is the proposed HRNet.

The input RGB image and output spectral images were randomly cropped to a $256 \times 256$ region, then rescaled to [0, 1]. The parameters of network are Xavier initialized~\cite{OrangeCat-Xavier}. The whole system was trained for 10000 epochs in total. The initial learning rate was $1\times 10^{-4}$ and halved every 3000 epochs. For optimization, the Adam optimizer was used with $\beta_1 = 0.5$ , $\beta_2 = 0.999$ and a batch size of 8. Reflection padding was used in the system to avoid border artifacts. The LeakyReLU activation~\cite{OrangeCat-lrelu} function was attached to each convolution layer. No normalization were used in the proposed architecture. All the experiments were performed on 2 NVIDIA Titan Xp GPUs.





\subsection{AIDAR - Cross-scale Aggregation Network for Spectral Reconstruction}        

 A cross-scale aggregation network (CSAN) is proposed with a novel feature fusion mechanism across multiple resolution branches.
In CSAN, the scale-wise residual dense groups (SRDGs) exploit hierarchical feature information over different spatial resolutions in parallel. 
The SRDG is a series of residual dense blocks to fully achieve all the hierarchical representation capability.
Here, a novel multi-scale feature fusion module is designed, which are named as cross-scale aggregation module (CSA), for compounding and aggregating the multi-resolution feature information from the prior SRDGs.
It generates refined features at each resolution-level by fusing useful information across all the scale-levels.
Such a function combines complementary characteristics from dynamic cross-scale representations in a stage-by-stage fashion.
Also, the hierarchical levels are extended to explore strong contextual information from the low-resolution representations.
It further has inner shortcut connections at each spatial level to improve gradient flow throughout the network.
In addition, a global skip connection routes data between two ends of the network, improving further the ability of the network to accurately recover ﬁne details.
Finally, the reconstruction block select useful set of features from each branch representations with step-wise refinement.
The CSAN can generate high-quality spectral images without noticeable artifacts, as will be confirmed by our results.

\begin{figure}[t]
\centering
    \includegraphics[width=\linewidth]{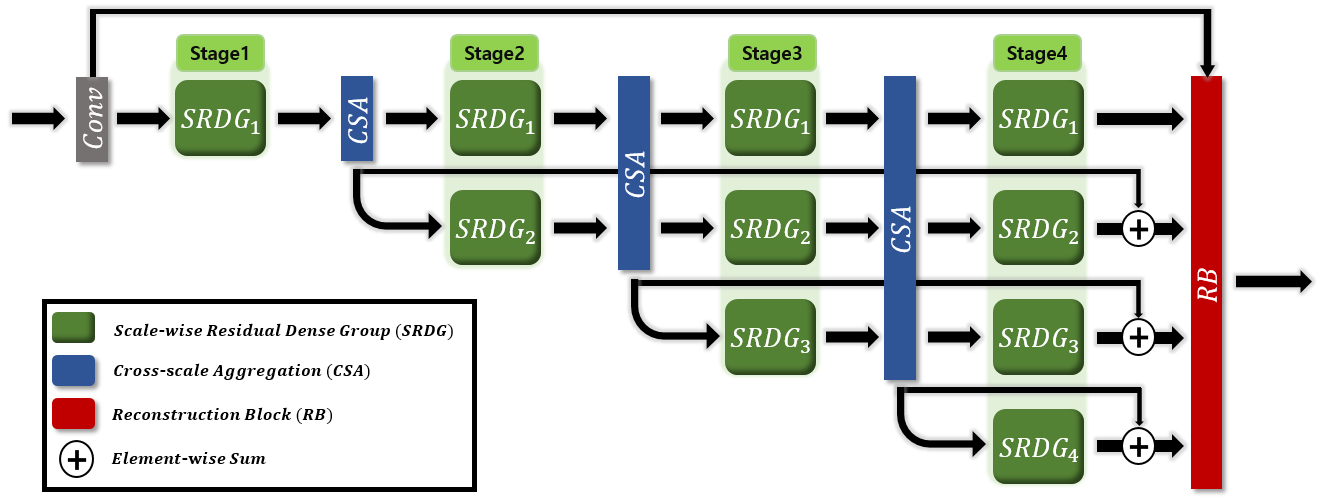}
    \caption{Overall framework of the AIDAR cross-scale aggregation network (CSAN).}\label{PARASITE_fig1}
\end{figure} 

\begin{figure}[t]
\centering
    \includegraphics[width=0.85\linewidth]{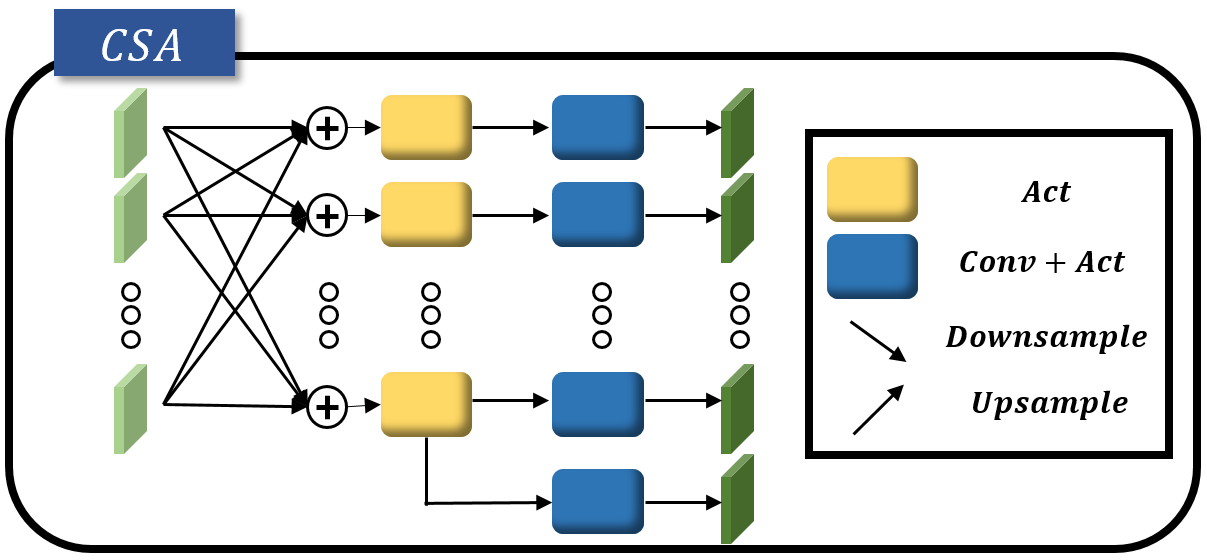}
    \caption{Cross-scale aggregation (CSA).}\label{PARASITE_fig2}
\end{figure} 

\subsubsection{Global Method Description}

Total method complexity: a pre-processing conv-block, 10 SRDGs(4 stages), a reconstruction block.

\noindent\textbf{Training}
training a CSAN roughly takes 16 hours with a single NVIDIA Titan XP GPU for 1000 epochs. Training input was Self-ensemble and model-ensemble, learning rate was 3e-5, batch size was 8, and the Adam optimizer was used.

\noindent\textbf{Testing}
testing a CSAN roughly takes about 30 seconds with a single Titan XP GPU. 

\subsubsection{Ensembles and fusion strategies}

Self-ensemble and model-ensemble were used. Quantitatively, MRAE of baseline model with the ensemble is about 0.015 higher than that without the ensemble. Moreover, it shows qualitative improvement with some great recovery of words in the hyperspectral images.

The baseline model is described in Figure~\ref{PARASITE_fig1}. For model ensemble, the model is modified by applying different width(the number of channels) and height(the number of stages) of the model, maintaining the volume of the models.


\subsection{VIPLab}
We separately learn the mean and corresponding residual of each image
using DNNs for the res learning net we normalize output feature to zero-mean.

\subsubsection{Ensembles and fusion strategies}
we train 3 network one uses RELU activation the other uses swish activation
finally a network trained with fine-tune result.


\subsection{TIC-RC - Hyperspectral Spectral Super-resolution via an Improved HSCNN+}
In the NTIRE2018 Spectral Reconstruction Challenge~\cite{arad2018ntire}, the HSCNN+ \cite{TIC-RC_HSCNN+} has achieved the best performance with a ResNet-based and a DenseNet-based approaches. Therefore, the ResNet-based HSCNN+ has been utilized as the baseline
, which has also shown a good performance on this challenge. However, there are two problems for the baseline method. The first one is the huge computational burden for us, one training epoch costs about 150s with our computational resource (Google Colab with K40 GPU), which is not acceptable. The second one is the reconstruction performance can be improved. According to that, an efficient framework with less ResNet blocks is desired whilst improving or maintaining the reconstruction performance. 

\begin{figure}[htbp]
    \centering
	\begin{subfigure}{0.15\linewidth}
		\includegraphics[width=\linewidth]{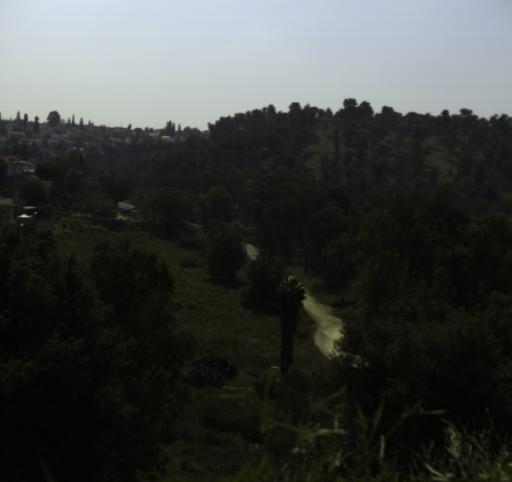}
		\caption{}
		\label{TIC-RC_fig:a}
	\end{subfigure}
	\begin{subfigure}{0.15\linewidth}
		\includegraphics[width=\linewidth]{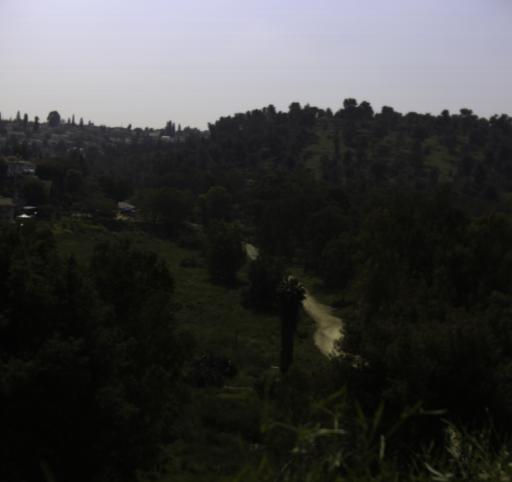}
		\caption{}
		\label{TIC-RC_fig:b}
	\end{subfigure}
	\begin{subfigure}{0.15\linewidth}
		\includegraphics[width=\linewidth]{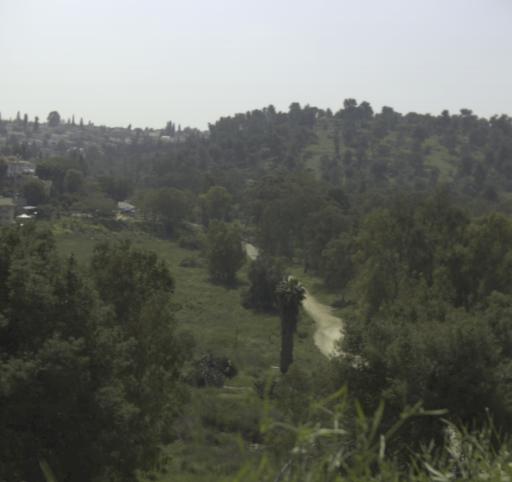}
		\caption{}
		\label{TIC-RC_fig:c}
	\end{subfigure}
	\begin{subfigure}{0.15\linewidth}
		\includegraphics[width=\linewidth]{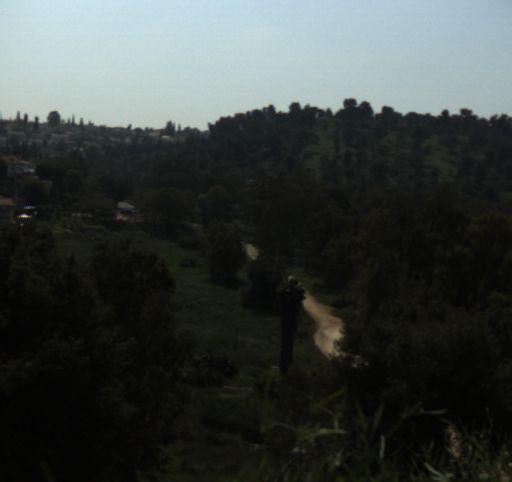}
		\caption{}
		\label{TIC-RC_fig:d}
	\end{subfigure}
	\begin{subfigure}{0.15\linewidth}
		\includegraphics[width=\linewidth]{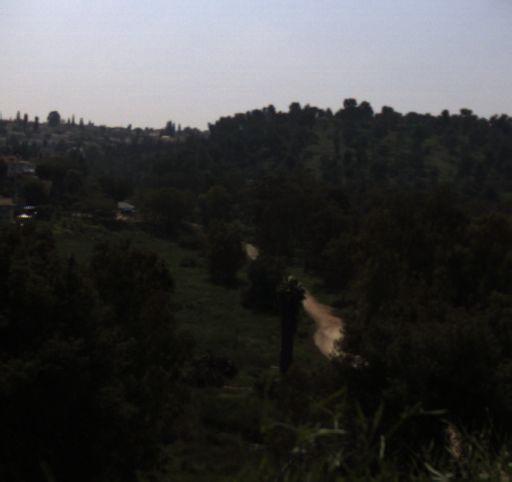}
		\caption{}
		\label{TIC-RC_fig:e}
	\end{subfigure}
	\begin{subfigure}{0.15\linewidth}
		\includegraphics[width=\linewidth]{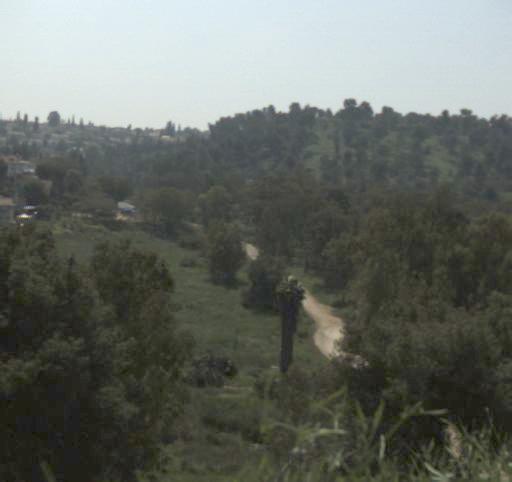}
		\caption{}
		\label{TIC-RC_fig:f}
	\end{subfigure}
	\caption{Comparison between original RGB images, generated WB images, and gamma images: (a) Clean image (b) WB of clean image (c) gamma of clean image, (d) real world image (e) WB of real world image, (f) gamma of real world image. }
	\label{TIC-RC_fig}
\end{figure}

\begin{figure}[htbp]
    \centering
	\begin{subfigure}{0.25\linewidth}
		\includegraphics[width=\linewidth]{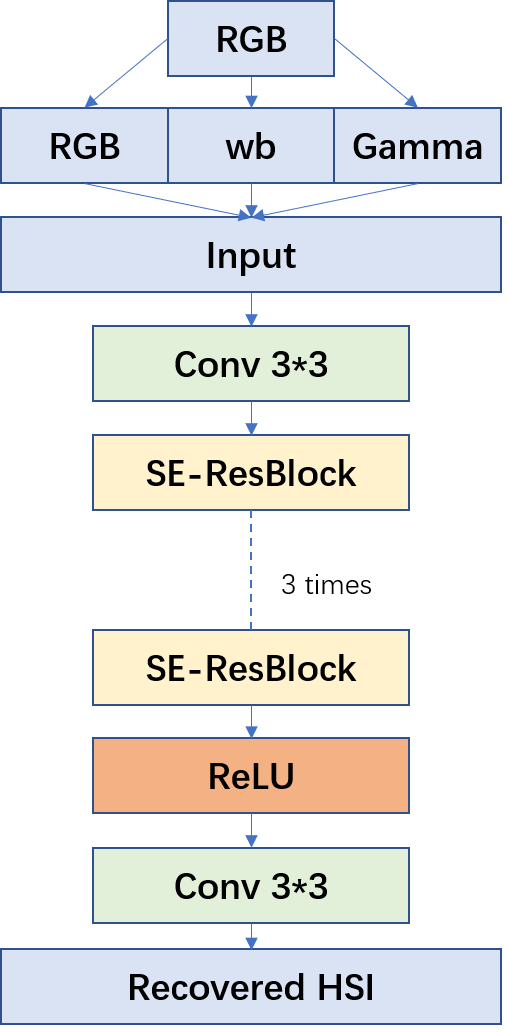}
		\caption{}
		\label{TIC-RC_fig:a2}
	\end{subfigure}
    ~~~~~~~~~~~~
	\begin{subfigure}{0.25\linewidth}
		\includegraphics[width=\linewidth]{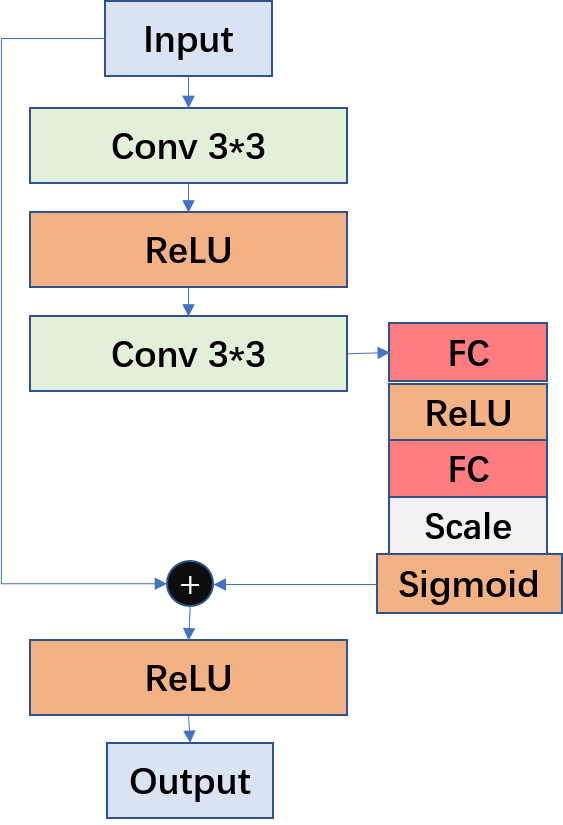}
		\caption{}
		\label{TIC-RC_fig:b2}
	\end{subfigure}
	\caption{(a) Improved HSCNN+ network framework (b) SE-ResBlock. }
	\label{TIC-RC_fig2}
\end{figure}

In the image dehazing field \cite{TIC-RC_REN}, multiple input RGB images after pre-processing have proved to be useful for recovering more information. Therefore, more RGB inputs are generated from the provided RGB images, including the white balanced (WB) image and the gamma correction (gamma) image. An example is shown in Figure \ref{TIC-RC_fig}. Figure~\ref{TIC-RC_fig2} depicts the suggested architecture. 

\begin{table}[ht]
	\begin{center}
	\resizebox{\linewidth}{!}
    {
		\begin{tabular}{|c|c|c|}
			\hline
			Methods & MRAE & Time(per epoch) \\
			\hline
			ResNet-based HSCNN+ &  0.0718251689 & 152s\\
			\hline
			Ours with raw input &  0.0714754468	 & 49s\\
			\hline
			Ours & 0.0687300712 & 50s\\
			\hline
		\end{tabular}
		}
		
		\caption{Track 2 reconstruction performance for baseline HSCNN+ and the proposed Improved HSCNN+ network.}
		\label{TIC-RC_table}
	
	\end{center}
\end{table}

\subsection{GD322 - Residual pixel attention network for spectral reconstruction from RGB Images\cite{Peng_2020_CVPR_Workshops}}
\begin{figure}[t]
\centering
    \includegraphics[width=1.0\linewidth]{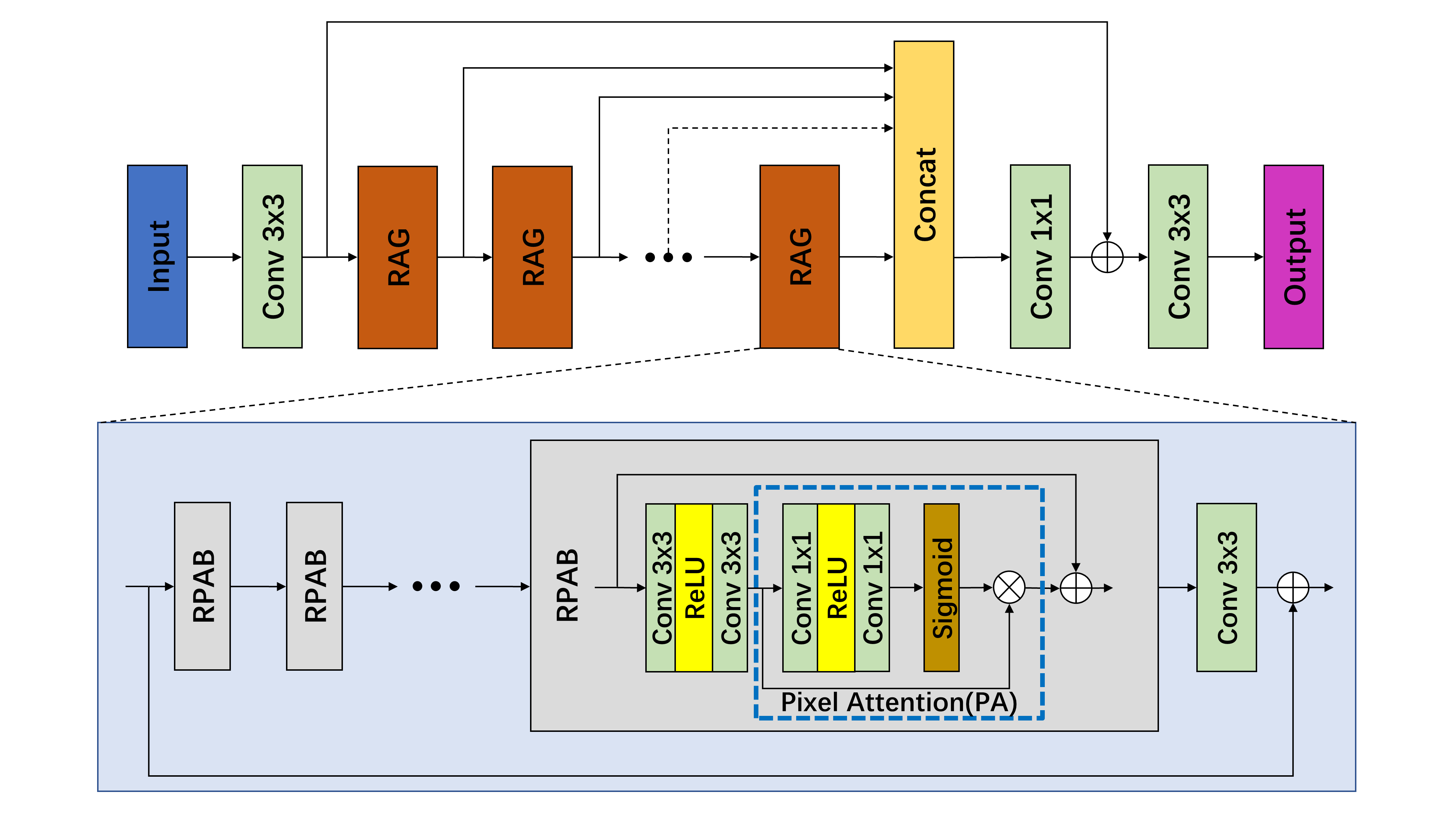}
    \caption{RPAN (up) and a RAG block (down).}
    \label{[GD322]_[RPAN]}
\end{figure}
A residual pixel attention network (RPAN) is designed for spectral reconstruction from RGB images as shown in Figure \ref{[GD322]_[RPAN]}.The RPAN we proposed was inspired by the RCAN\cite{zhang2018image}, so it should be noted that the proposed RPAN and RCAN have some similarities in composition.

The proposed network adopts a global residual architecture, then the main architecture of the RPAN network is constituted of $4$ residual attention group blocks (RAG) with $64$ filters. The RAGs also are stacked to form a Concat layer followed by $1\times1$ convolutions layer. By skip connection, they add $64$ feature maps from the first $3\times3$  convolution layer by global skip connection. Then output $31$ channel spectral image is through the last $3\times3$ convolution layer. Each RAG is composed of 8 RPABs (residual pixel attention block) with 64 filters, and the RPAB is composed of residual blocks and a novel module is called pixel attention (PA) inside the residual block.

The features of different positions and different channels should not be the same in importance, some positions and channels are more helpful for the spectral reconstruction. In order to better treat the features of different channels and different positions, the PA is firstly developed in our RPAN network, which can rescale the pixel-wise features in each channel adaptively to improve the quality of hyperspectral image reconstruction from RGB images, as shown in Figure \ref{[GD322]_[RPAN]}.

\noindent\textbf{Training}
During the training, the input RGB image and hyperspectral images are cropped into small pieces of $64\times64$ from the training dataset and the batchsize is set to $16$. In the “Clean” track, all biases are removed from each convolution layer, because the “Clean” track image contains no noise. In the “Real World” track, biases in all convolutional layers are reserved to compensate for the noise and JPG compression.
 
Adam optimizer is used for optimizing the proposed RPAN network with ${\beta}_1 = 0.9$, ${\beta}_2 = 0.999$, $\epsilon = 10^{-8}$ and the weight decay was set to $10^{-6}$. The initial learning rate is set to $8\times10^{-5}$, the learning rate decays by $0.8$ after every $5$ epochs, the network is ended the optimization at the 50-th epoch. MRAE loss function is used for training the RPAN network, zero-padding is used in all $3\times3$ convolutional layers to keep the feature map size unchanged. The proposed RPAN is trained by Pytorch platform on a single NVIDIA Titan Xp GPU. It takes about $13$ hours to train the RPAN network for each track.

\noindent\textbf{Testing}
Instead of cropping image into small blocks, the complete RGB image is used to get a complete spectral image on an NVIDIA Titan Xp. The RPAN network takes 1.35s(including inference time and spectral image reconstruction time) per image.


\subsection{LFB - Linear Spectral Estimate Refinement for Spectral Reconstruction from RGB}
The basic idea of the method used is that signal interpolation is more convenient when the underlying Laplacian signal energy is low.
To reduce the Laplacian signal energy, a linear estimation of the spectral stimulus is considered.
The proposed method can thus be described as two processing steps:
\begin{itemize}
  \item a direct linear estimation on the spectral stimulus based on known camera response functions and an appropriate spectral basis.
  \item a refinement of the initial estimate through a convolutional neural network.
\end{itemize}
The concept of refining an estimate on spectral signals is not new. 
The major difference of our approach is that the algorithm for obtaining the spectral estimate is handcrafted, explicitly dependant on the camera response and not subject to parameter optimization during network training.
The hybrid approach is therefore limited to the ``clean'' track.
\\
Based on the spectral estimate, the neural network is tasked with signal refinement.
Since the architecture of the convolutional neural network was not the focus of this work, a ResNet-18 was utilized.
All the weights were initialized using fixed-update initialization~\cite{lfb_zhang2018}.
The precise network architecture is summarized in Fig.~\ref{fig:lfb_resnet}.
\\
Only the new NTIRE2020 data was used for training and evaluation.
The network was trained using Adam optimization with an initial learning rate of $10^{-4}$ and both a patch size and batch size of 50.
The training itself was executed on a NVIDIA 2080TI graphics card.
Pytorch was utilized as a framework.
\\
It was found that the proposed hybrid approach outperforms the stand-alone ResNet and it was concluded that the hybrid approach is superior to the stand-alone ResNet. Figure~\ref{LBF_FIG} illustrates the suggested architecture and relative basis functions.
\tikzset{%
  >={Latex[width=2mm,length=2mm]},
  base/.style = {rectangle, draw=black,
                  minimum width=2cm, minimum height=0.3cm,
                  text centered},
  processing/.style = {base, rounded corners,},
  coord/.style = {coordinate},
  sum/.style      = {draw, rounded corners, circle, minimum width=0.1cm},
  relu/.style = {base, rounded corners, minimum width=1.5cm},
  resblock/.style = {base, rounded corners, minimum width=2cm, minimum height=0.5cm, fill=orange!15},
  rgbSpec/.style = {base, fill=gray!15},
}
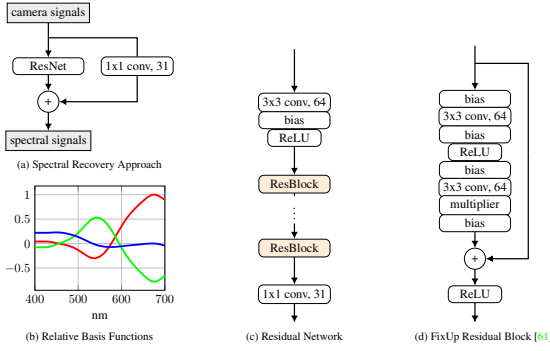
\begin{figure}[t]
    \resizebox{\linewidth}{!} {
    \centering
    \parbox{\textwidth}{
        \parbox[b]{0.33\textwidth}{
            \centering
            \begin{subfigure}[b]{0.3\textwidth}
            \centering
            \begin{tikzpicture}[node distance=0.5cm, every node/.style={fill=white}, align=center]
                \node (input)             [rgbSpec]                                           {camera signals};
                \node (help_in)           [coord, below of=input, yshift=-0.2cm] {};
                \node (resnet)            [processing, below of=input, yshift=-1cm]           {ResNet};
                \node (sum)               [sum, below of=resnet, yshift=-0.5cm]               {+};
                \node (output)            [rgbSpec, below of=sum, yshift=-0.6cm]              {spectral signals};
                \node (de)                [processing, right of=resnet, xshift=2.cm]          {1x1 conv, 31};
                
                \draw[->, line width=1pt](input) -- (resnet);
                \draw[line width=1pt]    (help_in) -| (de);
                \draw[->, line width=1pt](resnet) -- (sum);
                \draw[->, line width=1pt](de) |- (sum);
                \draw[->, line width=1pt](sum) -- (output);
            \end{tikzpicture}
            \caption{Spectral Recovery Approach}
            \label{fig:spec_recovery}
            \end{subfigure}
            \vskip1em
            \begin{subfigure}[b]{0.30\textwidth}
            \centering
            \pgfplotstableread[col sep=comma]{figures/linear_est.csv}\mydata
            \pgfplotsset{ytick style={draw=none}}
            \begin{tikzpicture}[align=center]
            \begin{axis}[
                width=\linewidth,
                xlabel = nm,
                grid=major,
                xmin = 400, xmax = 700,
            ]
                \addplot[red, line width=1.5pt] table[x index = {0}, y index = {1}]{\mydata};
                \addplot[green, line width=1.5pt] table[x index = {0}, y index = {2}]{\mydata};
                \addplot[blue, line width=1.5pt] table[x index = {0}, y index = {3}]{\mydata};
            \end{axis}
            \end{tikzpicture}
            \caption{Relative Basis Functions}
            \label{fig:spec_basis}
            \end{subfigure}
        }
      \hskip1em
      \begin{subfigure}[b]{0.28\textwidth}
        \centering
          \begin{tikzpicture}[node distance=0.5cm, every node/.style={fill=white}, align=center]
            \node (input)             [coord] {};
            \node (help_in)           [coord, below of=input] {};
            \node (conv1)             [processing, below of=input, yshift=-1cm] {3x3 conv, 64};
            \node (bias1)             [processing, below of=conv1]            {bias};
            \node (relu)              [relu, below of=bias1]                  {ReLU};
            \node (resblock1)         [resblock, below of=relu, yshift=-0.8cm]                   {ResBlock};
            \node (resblock2)         [resblock, below of=resblock1, yshift=-1.3cm]              {ResBlock};
            \node (vdots)             at ($(resblock1)!.4!(resblock2)$) {\vdots};
            \node (conv2)             [processing, below of=resblock2, yshift=-0.8cm]        {1x1 conv, 31};
            \node (output)            [coord, below of=conv2, yshift=-0.3cm] {ReLU};
              
            \draw[->, line width=1pt](input) -- (conv1);
            \draw[->,line width=1pt](relu) -- (resblock1);
            \draw[line width=1pt] (resblock1) -- (vdots);
            \draw[->,line width=1pt](vdots) -- (resblock2);
            \draw[->,line width=1pt](resblock2) -- (conv2);
            \draw[->,line width=1pt](conv2) -- (output);
          \end{tikzpicture}
        \caption{Residual Network}
        \label{fig:lfb_resnet}
      \end{subfigure}
      \hskip1em
      \begin{subfigure}[b]{0.28\textwidth}
        \centering
          \begin{tikzpicture}[node distance=0.5cm, every node/.style={fill=white}, align=center]
            \node (input)             [coord] {};
            \node (help_in)           [coord, below of=input] {};
            \node (bias1)             [processing, below of=input, yshift=-1cm]            {bias};
            \node (conv1)             [processing, below of=bias1]            {3x3 conv, 64};
            \node (bias2)             [processing, below of=conv1]            {bias};
            \node (relu)              [relu, below of=bias2]                  {ReLU};
            \node (bias3)             [processing, below of=relu]             {bias};
            \node (conv2)             [processing, below of=bias3]            {3x3 conv, 64};
            \node (mult)              [processing, below of=conv2]            {multiplier};
            \node (bias4)             [processing, below of=mult]             {bias};
            \node (sum)               [sum, below of=bias4, yshift=-0.5cm]    {+};
            \node (relu_f)            [relu, below of=bias4, yshift=-1.5cm] {ReLU};
            \node (output)            [coord, below of=relu_f, yshift=-0.3cm] {ReLU};
            \node (help)              [coord, right of=relu, xshift=1.cm] {};
              
            \draw[->,line width=1pt](input) -- (bias1);
            \draw[line width=1pt]    (help_in) -| (help);
            \draw[->,line width=1pt](bias4) -- (sum);
            \draw[->,line width=1pt](help) |- (sum);
            \draw[->,line width=1pt](sum) -- (relu_f);
            \draw[->,line width=1pt](relu_f) -- (output);
          \end{tikzpicture}
        \caption{FixUp Residual Block \cite{lfb_zhang2018}}
      \end{subfigure}
    }
    }
  \caption{Visualization of the LFB hybrid approach for spectral signal recovery from RGB.}
  \label{LBF_FIG}
\end{figure}


\subsection{CI Lab - HSCNND++: Dense-connection based CNN network for Hyperspectral reconstruction}

The backbone of this network is the DenseNet based model proposed in HSCNN+ \cite{TIC-RC_HSCNN+} that has shown superior performance for hyperspectral reconstruction during the NTIRE2018 Spectral Reconstruction Challenge. 

The backbone CNN model contains $2$ parallel branches, each containing $30$ dense-blocks with $4$ convolutional layers and a fusion block, which also has $30$ dense-blocks and merges the features from the two-parallel blocks and provids the predicted hyperspectral image. It was observed that the original HSCNN-D based model often got stuck in a local minima when trained with the mean relative absolute error(MRAE) or mean squared error(MSE), hence a compound loss function was proposed that achieved better convergence on the modified version of HSCNN-D model. This modified HSCNN-D model with the compound loss function is termed as HSCNND++. 

The compound loss function consists of three components: an absolute difference or $L1$ loss, a structural loss and a gradient loss. The $L1$ loss is  defined as $L_1 = ||G(x_{RGB}) - y_{HS}||_1$, which is a global loss that aims to reduce pixel-wise difference between the output, $G(x_{RGB})$, and the ground-truth, $y_{HS}$, images. To enforce further constraints, the structural and gradients losses were used. The structural loss is defined as $L_{SSIM} = 1 - SSIM(G(x_{RGB}), y_{HS})$ and the gradient loss is defined as $L_{\nabla} = ||\nabla(G(x_{RGB})) - \nabla(y_{HS}) ||_1$, where $\nabla$ is the Laplacian operator. The weights for the individual losses were $10,1,1$ for $L_1$, $L_{SSIM}$ and $L_\nabla$, respectively. 

\noindent\textbf{Training} The ``Clean'' model was trained for $20k$ epochs and the ``Real'' model for $10k$ epochs with a varying learning rate. The dataset was augmented by taking random crops of size $48\times48$. Random flips and rotation were also performed on the images at each iteration. Adam optimizer was used with $\beta_1=0.9$, $\beta_2=0.99$ for training the models. 

\noindent\textbf{Testing} guy
While testing, the model takes around $40ms$ to reconstruct the hyperspectral image from an input RGB image of dimensions $482\times512$, on a single NVIDIA 2080Ti GPU. 


\subsection{StaffsCVL - RGB to Spectral Reconstruction via Learned Basis Functions and Weights~\cite{Fubara_2020_CVPR_Workshops}}.

\begin{figure}[ht]
    \centering
    \includegraphics[width=\linewidth]{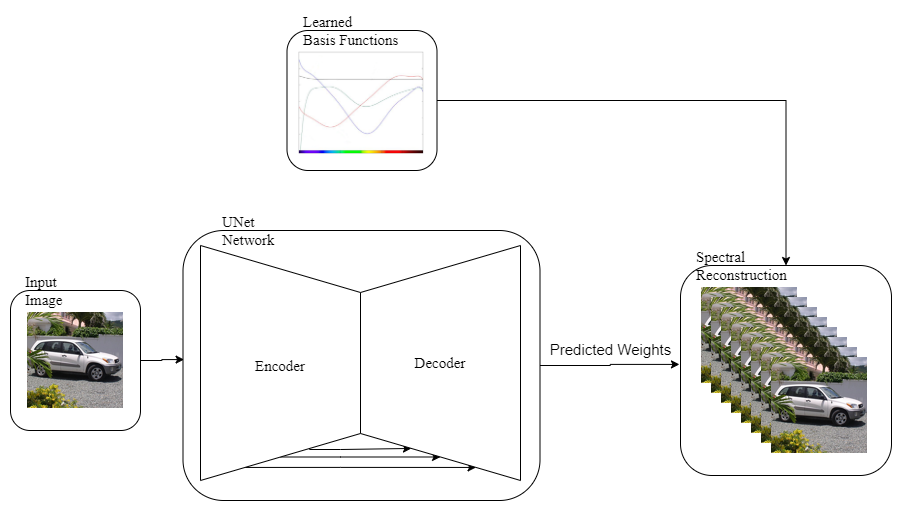}
    \caption{Architecture of the StaffsCVL method for spectral recovery.}
    \label{staffsCVL_fig:arch}
\end{figure}

Instead of predicting a 31-channel spectral image, the model predicts weights for a set of basis functions which are learned at the same time as the weights. In classical spectral reconstruction literature, the spectrum was recovered by weighted combination of basis functions\cite{agahian2008reconstruction} or sparse coding\cite{parmar2008spatio}. This method combines the simplicity of weighted basis functions and the performance and robustness of deep learning. The network predicts 10 weights for each pixel as well as learns a set of 10 basis functions which is then combined to form the final spectral image cube. \\
A modified UNet\cite{ronneberger2015u} network is used with skip connections to allow lower level features to flow to deeper layers. The 2x2 pooling layers are replaced with linear downsampling layers and four contracting steps are done. The cropping step before concatenation in the expansive path is replaced with a direct concatenation as cropping might dispose of edge information which could be useful for robust prediction, especially around the edges of the image. The same network architecture and training policy is used for both the clean and real world tracks. The proposed method is depicted in figure \ref{staffsCVL_fig:arch}.

The models from both tracks were trained on only the NTIRE 2020 challenge dataset without the use of any pre-trained models. They were trained on patches from the RGB image and spectral cubes. The RGB image and spectral cubes were resized to 512$\times$512, and 64$\times$64 patches were extracted deterministically, which were used for training. The training batch size was 128, learning rate was 1e-4 and the Adam optimiser~\cite{kingma2014adam} was used during training. Random horizontal and vertical flips were used for data augmentation, with a weight decay of 1e-5. The use of weight decay proved helpful to avoid overfitting. The basis functions are learned as a 10$\times$31 matrix variable during training without going through any neural network layer. During inference, the saved trained matrix is simply loaded into memory and used. At test time, the full RGB image is passed through the CNN. The spectral cube is then generated as a weighted combination of the basis functions, using the predicted weights. 
The proposed method is advantageous because it is able to reconstruct the spectral cube using fewer parameters than would normally be required (\ie predicting 10 weights per pixel instead of 31, a 67.74\% reduction in predicted output). This becomes even more significant when predicting 301 spectral bands (96.68\% reduction in this case). More detailed information on the method can be found in \cite{Fubara_2020_CVPR_Workshops}.


\subsection{Pixxel AI - MXR-U-Nets for Real Time Hyperspectral Reconstruction \cite{pixxel-ai-paper}}

The approach combines some of the very recent advancements in image classification, segmentation and Generative Adversarial Networks. At a high level, a model based on the U-Net~\cite{ronneberger2015u} architecture with self-attention \cite{pixxel-SAGAN} to project RGB images to their Hyperspectral counterparts was used. The loss function for the model is a slightly modified version of perceptual losses (See section 3.2 in \cite{pixxel-ai-paper}) \cite{pixxel-Perceptual} \cite{pixxel-StyleTransfer}.

\begin{figure}[ht]
    \centering
    \includegraphics[width=\linewidth]{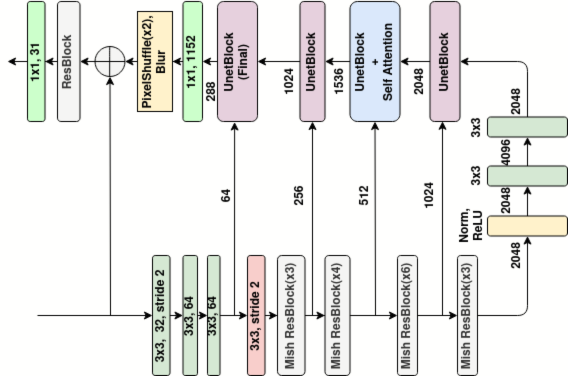}
    \caption{Architecture of the proposed mxresnet50 model.}
    \label{pixxel-ai:arch}
\end{figure}

\begin{figure}[ht]
    \centering
    \includegraphics[width=0.6\linewidth]{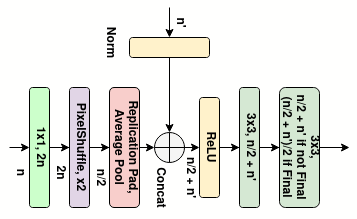}
    \caption{U-Net block inside the proposed mxresnet50 model.}
    \label{pixxel-ai:unet-block}
\end{figure}

The encoder backbone is an XResnet \cite{pixxel-BagOfTricks} with Mish \cite{pixxel-Mish} activation function (Replacing ReLU). The XResnet encoder, as proposed by He~\etal (referred to as Resnet-D in \cite{pixxel-BagOfTricks}), improves upon the original Resnet \cite{OrangeCat-ResNet} in classification performance. The solution combines both of these improvements and the model will be referred to as \textit{mxresnet}. \textit{mxresnet} implementation used in the implementation: \cite{pixxel-mxresnet}. Sub-pixel convolution layers \cite{OrangeCat-PixelShuffle} layers are used as upsampling layers in decoder blocks with ICNR \cite{pixxel-ICNR} initialization scheme and weight normalization \cite{pixxel-WeightNorm}. The sub-pixel convolution layers serve to conserve information during the upsampling part of a decoder. To reduce the checkerboard artifacts introduced by the sub-pixel convolution layers, ICNR initialization with weight normalization is used. Each sub-pixel convolution layer is followed by a blur \cite{pixxel-Blur} layer consisting of an average pooling layer with a $2\times2$ filter and a stride of $1$. This improvement adds to the previous solution for dealing with checkerboard artifacts in the outputs of pixel shuffle layers. The decoder has a Self Attention block as proposed by Zhang~\etal in \cite{pixxel-SAGAN} to help the network focus more on the relevant parts of the image. Figure~\ref{pixxel-ai:arch} and Figure~\ref{pixxel-ai:unet-block} depict the proposed architecture and U-Net block respectively.\\

\noindent\textbf{Training} An \textit{mxresnet50} encoder based model is trained for $200$ epochs with the AdamW \cite{pixxel-adamw} optimizer with a weight decay of $1e-3$. The learning rate follows the OneCycle schedule \cite{pixxel-OneCycle}. Under this schedule, the learning rate is started at 1e-5 and increased to $1e-3$ over $60$ epochs following a half cosine curve. After the learning rate peaks, it is reduced to $1e-9$ over another $140$ epochs following a similar half cosine curve. The model was trained using mixed-precision training \cite{pixxel-MixedPrecision} to lower training time and memory requirements. A single V100 GPU was used for all training runs.


\subsection{Image Lab - Light Weight Residual Dense Attention Net for Spectral Reconstruction}
\begin{figure}
  \centering
     \includegraphics[width=0.5\linewidth]{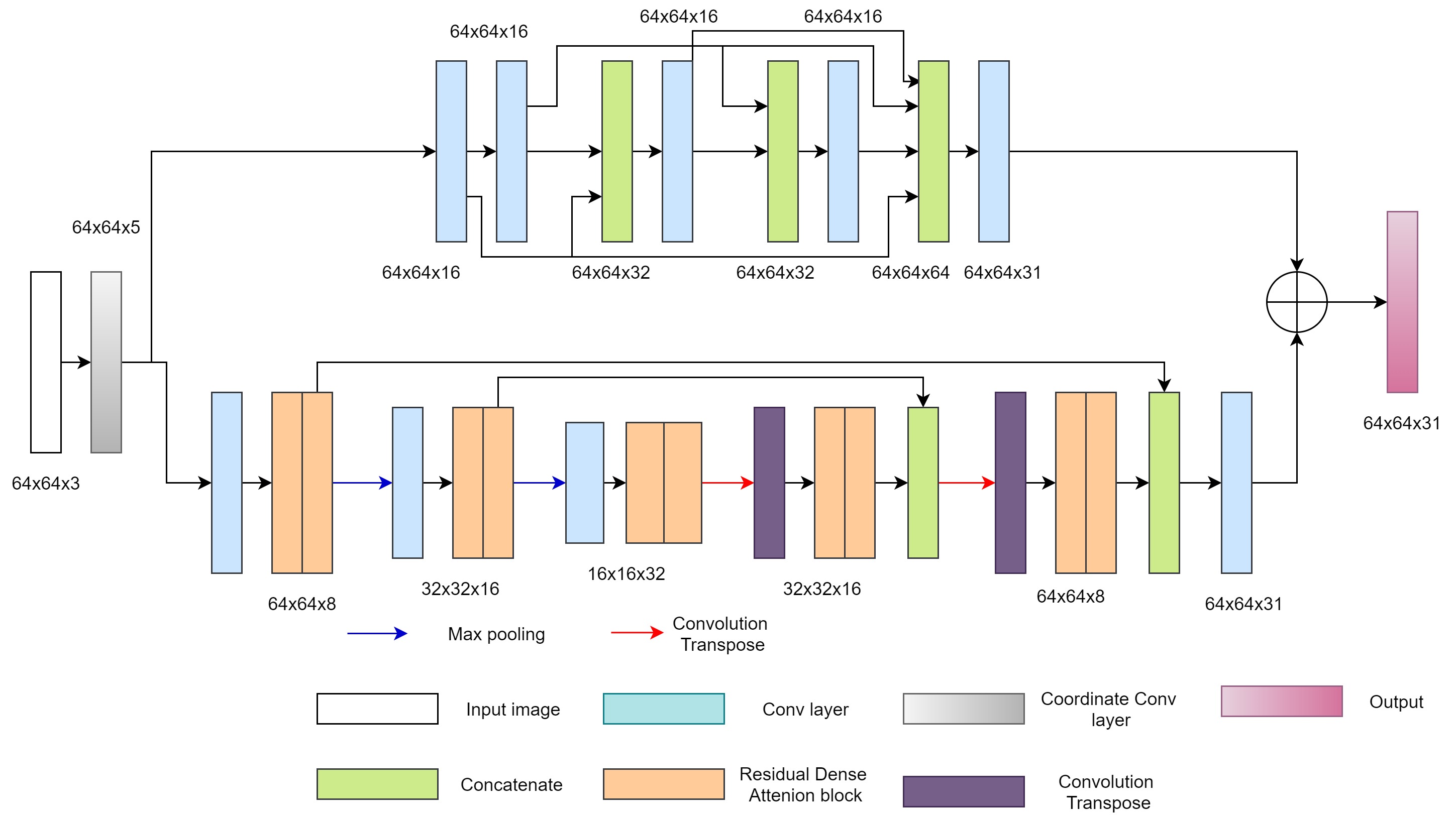}
  \caption{Architecture of the Light Weight Residual Dense Attention Net for Spectral Reconstruction.}
  \label{image_lab_fig_1}
\end{figure}

An ensemble of convolution layer with Residual Dense Attention block (RDAB) connected at multi-scale level are used for spectral reconstruction. Specifically, in each block, certain significant features are given more importance spatially and spectrally by its dedicated attention mechanism~\cite{ImageLab-CBAM}, henceforth multi-scale hierarchical features are extracted at multi-level to widen the forward paths for higher capacity. The proposed network ~\cite{ImageLab-LWRDA} for Spectral Reconstruction is shown in Figure~\ref{image_lab_fig_1}.

\begin{figure}
  \centering
     \includegraphics[width=0.5\linewidth]{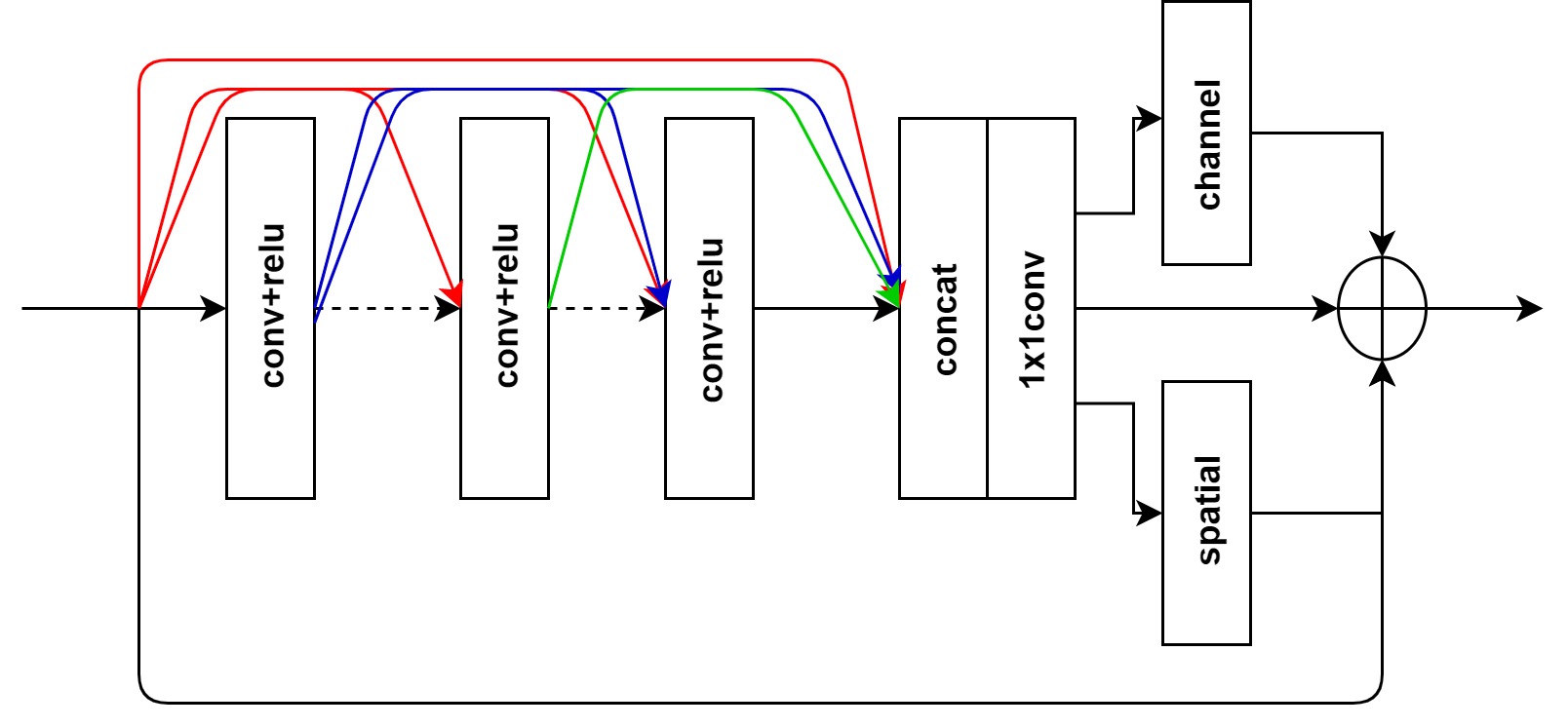}
  \caption{Residual Dense Attention Block.}
  \label{image_lab_fig_2}
\end{figure}

In this Network, the RGB image serves as an input, the coordinal features are extracted to improve its spatial information~\cite{ImageLab-coord}. The weights from the Coordinate convolution block are shared by two independent feature extraction mechanisms, one by dense feature extraction and the other by the multiscale hierarchical feature extraction. The Dense features are extracted by a dedicated dense block~\cite{ImageLab-Residualdense} connection whereas the multiscale hierarchical features are extracted by the Residual Dense Attention Block.  The block diagram of Residual Dense Attention Block is shown in Figure~\ref{image_lab_fig_2}. The Residual Dense Block (RDB) generates the local hierarchical feature. RDAB blocks are connected at multi-scale level in a U-net fashion, where the encoding phase consists of Maxpooling layer in between the RDAB blocks meanwhile the decoding phase consists of Transpose Convolution between them. This Transpose Convolution helps to reconstruct the image to the same spatial resolution as that of the input. Finally, the features from both the feature extraction mechanisms are globally fused to produce the 31 spectral bands.  

\[Loss Function=L2+(1-SSIM)\]

\section*{Acknowledgments}

We thank the NTIRE 2020 sponsors: HUAWEI Technologies Co. Ltd., OPPO Mobile Corp., Ltd., Voyage81, MediaTek Inc., DisneyResearch$\mid$Studios, and ETH Zurich (Computer Vision Lab). Graham Finlayson is grateful for the support of EPSRC grant EP S028730. Ohad Ben-Shahar gratefully acknowledges the support of the ISF-FIRST program grant 555/19. 

\appendix
\section{Teams and affiliations}
\label{sec:appendix}
\subsection*{NTIRE2018 team}
\noindent\textit{\textbf{Title: }} NTIRE 2018 Challenge on Spectral Reconstruction from RGB Images

\noindent\textit{\textbf{Members: }} \textit{Boaz Arad$^{1,4}$\noindent(boazar@post.bgu.ac.il)}, Radu Timofte $^{2}$, Ohad Ben-Shahar $^{1}$, Yi-Tun Lin $^{3}$, Graham Finlayson $^3$, Shai Givati $^1$

\noindent\textit{\textbf{Affiliations: }}\\
$^1$ Ben-Gurion University of the Negev, Israel\\
$^2$ Computer Vision Lab, ETH Zurich, Switzerland\\
$^3$ University of East Anglia, UK\\
$^4$ Voyage81
\vspace{-0.2cm}

\subsection*{IPIC\_SSR}
\noindent\textit{\textbf{Title: }} Adaptive Weighted Attention Network with Camera Spectral Sensitivity Prior for Spectral Reconstruction from RGB Images

\noindent\textit{\textbf{Members: }} \textit{Jiaojiao Li$^{1}$ (jjli@xidian.edu.cn)}, Chaoxiong Wu$^1$, Rui Song$^1$, Yunsong Li$^1$, Fei Liu$^1$\\
\noindent\textit{\textbf{Affiliations: }}\\
$^1$ Xidian University, Xi’an, China\\
\vspace{-0.2cm}

\subsection*{MDISL-lab}
\noindent\textit{\textbf{Title: }} Improved Pixel-wise Deep Function-Mixture Network

\noindent\textit{\textbf{Members: }} \textit{Zhiqiang Lang$^{1}$ (2015303107lang@mail.nwpu.edu.cn)}, Wei Wei$^1$, Lei Zhang$^1$, Jiangtao Nie$^1$

\noindent\textit{\textbf{Affiliations: }}\\
$^1$ Chang’an campus of Northwestern Polytechnical University\\
\vspace{-0.2cm}

\subsection*{OrangeCat}
\noindent\textit{\textbf{Title: }} Hierarchical Regression Network for Spectral Reconstruction from RGB Images

\noindent\textit{\textbf{Members: }} \textit{Yuzhi Zhao$^{1}$ (yzzhao2-c@my.cityu.edu.hk)}, Lai-Man Po$^{1}$, Qiong Yan$^{2}$, Wei Liu$^{2,3}$, Tingyu Lin$^{1}$

\noindent\textit{\textbf{Affiliations: }}\\
$^1$ City University of Hong Kong\\
$^2$ SenseTime Research\\
$^3$ Harbin Institute of Technology\\
\vspace{-0.2cm}

\subsection*{AIDAR}
\noindent\textit{\textbf{Title: }} Cross-scale Aggregation Network for Spectral Reconstruction

\noindent\textit{\textbf{Members: }} \textit{Youngjung Kim$^{1}$ (read12300@add.re.kr)}, Changyeop Shin$^{1}$, Kyeongha Rho$^{1}$, Sungho Kim$^{1}$

\noindent\textit{\textbf{Affiliations: }}\\
$^1$ Agency for Defense Development\\
\vspace{-0.2cm}

\subsection*{VIPLab}
\noindent\textit{\textbf{Title: VIPLab}} 

\noindent\textit{\textbf{Members: }} \textit{Junhui HOU (zbzhzhy@gmail.com)}, Zhiyu ZHU, Yue QIAN

\noindent\textit{\textbf{Affiliation: }} City University of Hong Kong\\
\vspace{-0.2cm}

\subsection*{TIC-RC}
\noindent\textit{\textbf{Title: }} Hyperspectral Spectral Super-resolution via an Improved HSCNN+

\noindent\textit{\textbf{Members: }} \textit{He Sun$^{1}$ (h.sun@strath.ac.uk)}, Jinchang Ren$^1$, Zhenyu Fang$^1$, Yijun Yan$^1$

\noindent\textit{\textbf{Affiliations: }}\\
$^1$ University of Strathclyde, Glasgow, U.K.\\
\vspace{-0.2cm}

\subsection*{GD322}
\noindent\textit{\textbf{Title: }} Residual Pixel Attention Network for Spectral Reconstruction from RGB Images

\noindent\textit{\textbf{Members: }} \textit{Hao Peng$^{1}$ (Hao\_Peng@outlook.com)}, Xiaomei Chen$^1$, Jie Zhao$^1$

\noindent\textit{\textbf{Affiliations: }}\\
$^1$ Beijing Institute of Technology\\
\vspace{-0.2cm}

\subsection*{LFB}
\noindent\textit{\textbf{Title: }} Linear Spectral Estimate Refinement for Spectral Reconstruction from RGB

\noindent\textit{\textbf{Members: }} \textit{Tarek Stiebel$^{1}$ (tarek.stiebel@lfb.rwth-aachen.de)}, Simon Koppers$^1$, Dorit Merhof$^1$

\noindent\textit{\textbf{Affiliations: }}\\
$^1$ Institute of Imaging \& Computer Vision, RWTH Aachen University\\
\vspace{-0.2cm}

\subsection*{CI Lab}
\noindent\textit{\textbf{Title: }} HSCNND++: Dense-connection based CNN network for Hyperspectral reconstruction

\noindent\textit{\textbf{Members: }} \textit{Honey Gupta$^{1}$ (hn.gpt1@gmail.com)}, Kaushik Mitra$^1$

\noindent\textit{\textbf{Affiliations: }}\\
$^1$ Indian Institute of Technology Madras, Chennai\\
\vspace{-0.2cm}

\subsection*{StaffsCVL}
\noindent\textit{\textbf{Title: }} RGB to Spectral Reconstruction via Learned Basis Functions and Weights

\noindent\textit{\textbf{Members: }} \textit{Biebele Joslyn Fubara$^{1}$ (fubarabjs@yahoo.co.uk)}, Dave Dyke$^1$, Mohamed Sedky$^1$

\noindent\textit{\textbf{Affiliations: }}\\
$^1$ Staffordshire University, UK\\
\vspace{-0.2cm}

\subsection*{Pixxel AI}
\noindent\textit{\textbf{Title: }} MXR-U-Nets for Real Time Hyperspectral Reconstruction

\noindent\textit{\textbf{Members: }} \textit{Akash Palrecha$^{1}$ (akashpalrecha@gmail.com)}, Atmadeep Banerjee$^{1}$

\noindent\textit{\textbf{Affiliations: }}\\
$^1$ {Pixxel} (\url{https://www.pixxel.space})\\
\vspace{-0.2cm}

\subsection*{Image Lab}
\noindent\textit{\textbf{Title: }} Light Weight Residual Dense Attention Net for Spectral Reconstruction

\noindent\textit{\textbf{Members: }} \textit{Sabarinathan$^{1}$ (sabarinathantce@gmail.com)}, K Uma$^2$, D Synthiya Vinothini$^2$, B Sathya Bama$^2$ ,S M Md Mansoor Roomi$^2$

\noindent\textit{\textbf{Affiliations: }}\\
$^1$ Couger Inc\\
$^2$ Thiagarajar college of engineering\\
\vspace{-0.2cm}




{\small
\bibliographystyle{ieee}
\bibliography{NTIRE2020_Spectral}
}

\end{document}